# THE TEMPORAL DEVELOPMENT OF DUST FORMATION AND DESTRUCTION IN NOVA SAGITTARII 2015#2 (V5668 SGR): A PANCHROMATIC STUDY

R. D. GEHRZ[1], A. EVANS[2], C. E. WOODWARD[1], L. A. HELTON[3], D. P. K. BANERJEE[4], M. K. SRIVASTAVA[4], N. M. ASHOK[4], V. JOSHI[4], S. P. S. EYRES[5], JOACHIM KRAUTTER[6], N. P. M. KUIN[7], K. L. PAGE[8], J.P. OSBORNE[8], G. J. SCHWARZ[9], D. P. SHENOY[1], S. N. SHORE[10], S. G. STARRFIELD[11], R. M. WAGNER[12]

[1] Minnesota Institute for Astrophysics, School of Physics and Astronomy, 116 Church Street, S. E., University of Minnesota, Minneapolis, Minnesota 55455, USA

gehrz@astro.umn.edu

[2] Astrophysics Group, Keele University, Keele, Staffordshire, ST5 5BG, UK

[3] USRA-SOFIA Science Center, NASA Ames Research Center, Moffett Field, CA 94035, USA

[4] Astronomy and Astrophysics Division, Physical Research Laboratory, Ahmedabad 380009, India

[5] Faculty of Computing, Engineering & Science, University of South Wales, Pontypridd CF37 1DS, United Kingdom; Jeremiah Horrocks Institute, University of Central Lancashire, Preston PR1 2HE, UK

[6] Landessternwarte, Zentrum für Astronomie der Universitaet Heidelberg, Koenigstuhl 12, D-69117 Heidelberg

[7] Mullard Space Science Laboratory, University College London, Holmbury St. Mary, Dorking, Surrey RH5 6NT, UK

[8] Department of Physics and Astronomy, University of Leicester, Leicester, LE1 7RH, UK

[9] American Astronomical Society, 1667 K Street, NW, Suite 800, Washington, DC 20006, USA

[10]Dipartimento di Fisica `Enrico Fermi', Università di Pisa, I-56127 Pisa, Italy; INFN-Sezione Pisa, Largo B. Pontecorvo 3, I-56127 Pisa, Italy

[11]School of Earth and Space Exploration, Arizona State University, Box 871404, Tempe, AZ 85287-1404, USA

[12]Department of Astronomy, The Ohio State University, 140 West 18th Avenue, Columbus, OH 43210, USA

ABSTRACT

We present 5-28 μm SOFIA FORECAST spectroscopy complemented by panchromatic X-ray through infrared observations of the CO nova V5668 Sgr documenting the formation and destruction of dust during ~500 days following outburst. Dust condensation commenced by 82 days after outburst at a temperature of ~1090 K. The condensation temperature indicates that the condensate was amorphous carbon. There was a gradual decrease of the grain size and dust mass during the recovery phase. Absolute parameter values given here are for an assumed distance of 1.2 kpc. We conclude that the maximum mass of dust produced was $1.2 \times 10^{-7}\,M_\odot$ if the dust was amorphous carbon. The average grain radius grew to a maximum of ~ 2.9 μm at a temperature of ~720 K around day 113 when the shell visual optical depth was $\tau_v$ ~ 5.4. Maximum grain growth was followed by followed by a period of grain destruction. X-rays were detected with Swift from day 95 to beyond day 500. The Swift X-ray count rate due to the hot white dwarf peaked around day 220, when its spectrum was that of a kT = 35 eV blackbody. The temperature, together with the super-soft X-ray turn-on and turn-off times, suggests a WD mass of ~ 1.1 $M_\odot$. We show that the X-ray fluence was sufficient to destroy the dust. Our data show that the post-dust event X-ray brightening is not due to dust destruction, which certainly occurred, as the dust is optically thin to X-rays.

1. INTRODUCTION

Gas and dust are injected into the interstellar medium (ISM) by classical nova (CN) explosions caused by thermonuclear runaways (TNRs) on the surfaces of white dwarf (WD) stars in close binary systems (see Bode & Evans 2008, Woudt & Riberio 2014 for overviews). Infrared (IR) observations of the temporal development of CN eruptions give important physical parameters that characterize the nova explosion and reveal the extent to which nova ejecta affect ISM abundances on both local and global scales (Gehrz 1988, 1999, 2008; 2012; Gehrz et al. 1998; Evans and Gehrz 2012; Gehrz, Evans and Woodward 2013). Nova explosions on carbon-oxygen (CO) WDs can result in the condensation of grains whose composition includes carbon, silicates, SiC, and hydrocarbons (see Evans & Rawlings 2008, Gehrz 2008, Evans & Gehrz 2012). Nova explosions on more massive oxygen-neon (ONe) WDs have produced ejecta with substantial enrichments of C, N, O, Ne, Mg, and Al over solar abundance (e.g. Helton et al. 2012, Woodward & Starrfield 2011). Several ONe novae have also produced dust.

Here we report Stratospheric Observatory for Infrared Astronomy (SOFIA; Young et al. 2012) Faint Object infraRed CAmera for the SOFIA Telescope (FORCAST; Herter et al. 2012) grism spectroscopy of an optically thick dust formation episode during the temporal development of the CO nova V5668 Sgr (Nova Sgr 2015 #2; PNV J18365700-2855420 ). Supporting observations include IR spectroscopy from Mt Abu, optical spectroscopy from the Michigan-Dartmouth-MIT (MDM), Multiple Mirror Telescope (MMT), and NASA Hubble Space Telescope (HST) observatories, and UV/X-ray Swift observations that span more than ~500 days beginning 2.5 days after the outburst.

Discovered on 2015 March 15.634 UT (JD 2457097.134, which we take to be day zero) by John Seach (Seach 2015), V5668 Sgr rose in visual brightness to $m_v$ ~+4.4 on March 21.6742 (JD 2457103.174; day 6).

Our Swift XRT observations show the distinct gradual rise and decline of both a hard and a super-soft X-ray component ascribed to shocks in the ejecta and the hot photosphere of the WD respectively. This behavior has been commonly seen in Swift campaigns on other novae (Osborne 2015). UVOT UV filter and grism spectra clearly delineate the strongly-dust-dominated evolution. This is the first nova exhibiting a strong dust formation event to have contemporary high-cadence Swift monitoring observations. Although Swift measured the X-ray flux of the dust-forming nova V2362 Cyg (Lynch et al. 2008), the observations were sporadic and insufficient to clarify the relationship between the temporal evolution of the dust emission and the X-rays. Dust formation can potentially hide metals from the X-ray flux in high optical depth grains, offering the possibility to explore the dust formation process via its effect on the absorption of soft X-rays.

## 2. OBSERVATIONS

### 2.1 SOFIA/FORCAST Grism Observations

Mid-IR spectroscopic observations of V5668 Sgr were obtained with FORCAST on SOFIA as previously described in summarizing our similar observations of nova V339 Del (Gehrz et al. 2015). The data were acquired during two flights: F216; originating from Palmdale, CA at an altitude of 43,000 ft, (13.11 km) on 2015 June 5 UT (JD 2,457,178.84; day 81.7) and F226; originating from Christchurch NZ at an altitude of 37,000 ft, (11.28 km) on 2015 July 6 UT (JD 2,457,209.88; day 112.75) as part of our SOFIA Cycle 3 Target-of-Opportunity program to observe CNe in outburst (P.I. Gehrz, Program ID 03_0020). The visual light curve and the timing of the observations with respect to the dust formation event are shown in Figures 1a and b. The FORCAST spectra are shown in Figure 2[1]. Two additional epochs of SOFIA FORCAST 5 – 27 μm spectra were obtained during the late stages of the dust evolution as part of our SOFIA Cycle 4 follow-up program (P.I. Helton, PID 04_0181). These include a spectrum obtained on 2016 July 16 UT (JD 2457585.5; day 488.4); a preliminary reduction is shown in Figure 2. These data are still under analysis and will be the subject of a forthcoming paper.

Acquisition images with FORCAST were taken using the F111 filter in the short-wave channel and were used to position the target on the slit. During the first epoch, spectroscopy was performed using all four FORCAST grisms, G063, G111, G227, and G329, with spectral coverage from 4.9 – 37.1 μm. Due to the low signal at long wavelengths, the long wavelength grism observations during the second visit were replaced with broad-band photometry.

A summary of the observations is provided in Tables 1 and 2. Imaging data were reduced using the SOFIA Data Reduction Pipeline for FORCAST, v1.0.6, and grism data with v1.0.5 (Clark et al. 2014). See Gehrz et al. (2015) for additional details of the overall observing strategy and data reduction employed.

## 2.2 Swift Observations

The Neil Gehrels Swift Observatory (Gehrels et al. 2004) began observations of V5668 Sgr on 2015 March 18 (JD 2457099.5), 2.4 days after the outburst. A one ks exposure was obtained approximately daily, with the occasional interruption caused by GRBs or high-priority ToO requests, until the end of 2015 March. Given the high optical brightness of the nova, the X-ray Telescope (XRT; Burrows et al. 2005) had to be operated in the 1-dimensional Windowed

---

[1] The reduced SOFIA FORCAST data are available in the SOFIA Science Data Archive at
https://dcs.sofia.usra.edu/dataRetrieval/SearchScienceArchiveInfo.jsp

Timing (WT) mode, to minimize the CCD read-out time and prevent the optical photons from building up sufficiently to appear as spurious X-ray events[2]. The UV/Optical Telescope (UVOT; Roming et al. 2005) was necessarily blocked for these early observations. WT mode is less sensitive to faint sources than is the 2-dimensional Photon Counting (PC) mode because of a higher background, and no X-ray source was detected during this interval. The cadence of observations was then decreased to once every three days throughout 2015 April, given the non-detection, and then to around once a week during May and until mid-June. At this time, the optical brightness had faded sufficiently for PC mode to be used safely, and for UV photometric and UV grism (170nm - 500nm) data to be collected by the UVOT. V-grism (275nm - 660nm) data could also be usefully collected during the dip in brightness from June 24 (JD 2457197.5; day 101) to August 3 (JD2457237.5; day141).

The Swift-UVOT grism spectra (170.0-660.0nm) were extracted using the UVOTPY software (Kuin 2014) using the calibration described in Kuin et al. (2015a). The UVOT grism spectra tend to have wavelength offsets. Therefore the wavelength scale of each spectrum was adjusted to match the HST spectrum using the emission line at 214.7nm or at 232.5nm. Contamination by the zeroth or first orders of other sources in the field was identified by inspection of the grism images. Areas of the spectrum that were affected by zeroth orders, or were too bright, have been removed. Second order lines are also present, most noticeable is the 175.0(2) nm NIII line which is seen at 294.4 nm in the first order spectrum, though in most cases the spectrum placement on the detector was such that the second order partly fell along the first order with reduced signature in the first order (see the discussion in Kuin et al. 2015a). The UVOT spectral observations started about midway during the drop in brightness from 6th to 12$^{th}$ magnitude. The drop in brightness was rapid enough that a significant effect was seen from day to day. The overall shape of the spectra shows a fairly even decrease in brightness throughout the ultraviolet, a behavior consistent with the neutral UV extinction expected for large grains. The Swift-UVOT photometry was performed using the UV filters. Due to the high brightness, the readout-streak of the source on the image was used to derive the magnitudes using the calibration and software of Page et al. (2013a). Only during the period July 3 – 20, 2015, was the nova faint enough for normal UVOT aperture photometry that was reduced using the calibration published by Breeveld et al. (2011).

---

[2] This is known as optical loading, See http://www.swift.ac.uk/analysis/xrt/optical_loading.php for more details.

On 2015 June 18 (JD 2457191.5; day 95), Swift-XRT first detected an X-ray source, at a level of 6.1 (+2.5/-1.9) x $10^{-3}$ count $s^{-1}$. A slow brightening of this source was then observed until around August 23 (JD 2457257.5; day 161; count rate ~ 0.06 count $s^{-1}$); after the next observation, on August 30 (JD 2457264.5; day 168; count rate ~ 0.56 count $s^{-1}$), the rate of brightening increased markedly, with the spectrum becoming significantly softer, reaching a peak of ~12 count $s^{-1}$ around October 25 (JD 2457320.5; day 224). Given the appearance of this super-soft emission (see Figures 3 and 4), daily observations of ~1.5 ks were performed when possible for four weeks starting from August 31 (JD 2457265.5; day 169). In addition, on September 25 (JD 2457290.5; day 193) a series of short snapshots were obtained every Swift 96-min orbit, to search for the high-amplitude rapid variability that has been seen in other novae (e.g., RS Oph - Osborne et al. 2011; KT Eri - Beardmore et al. 2010; V339 Del - Beardmore, Osborne & Page 2013). The data sets were also searched for shorter quasi-periodic oscillations, another type of variability sometimes found in the brighter novae (Ness et al. 2015). Data were then collected every 1-2 days until the nova became too close to the Sun for Swift to observe on November 13 (JD 2457339.5; day 243). During this interval, UVOT UV grism spectra and photometry were collected. After October 25 (JD 2457320.5; day 224), the X-ray light-curve showed a declining trend, reaching 3.87 +/- 0.05 count s$^{-1}$ just before entering the solar observing constraint. Once the nova had re-emerged from behind the Sun, a Swift observation was obtained, finding a much lower count rate of 0.077 +/- 0.007 count s$^{-1}$ on 2016 February 16 (JD 2457434.5; day 338). Data were then collected once a week until the end of April, and once a month until August (out to 506 days post discovery), showing a slow decline in the X-ray flux from the end of April onward, following a short plateau. A final observation was performed on 2017 July 10 (2457944.5; day 848), at which time the X-ray count rate was at a level of $(9 \pm 2) \times 10^{-3}$ count s$^{-1}$, consistent with the previous measurement on 2016 August 2. The UVOT UV spectra continued to show bright emission lines and a gradual decline in brightness.

With respect to the monitoring UVOT spectra (see Figure 5), strong emission lines were observed of N III 1750 Å, C III 1909 Å, N II 2143 Å, C II, 2324 Å, Mg II 2800 Å, the N III/C III 4630/4650 Bowen blend, [O III] 4959/5007 Å, and several H Balmer lines were seen as well. The Balmer continuum became more pronounced during the brightness minimum. Except during the minimum brightness the core of the strongest emission lines were too bright for calibration. The N III 1750 Å and the N II 2143 Å lines were the least affected. In Table 3 the average fluxes in 2 nm wide regions are listed and also the 175.0 NIII and 214.3 NII flux and FWHM from fitting a Gaussian to the line. The temporal development of the continuum and line fluxes given in Table 3 is shown in Figure 3.

The super-soft X-ray light-curve of V5668 Sgr shows high amplitude flux variability during its rise to peak flux, fading and re-brightening by up to an order of magnitude over 2-3 days (see Figs 3 and 4). This behavior appears similar to that seen in V959 Mon (Page et al. 2013), V339 Del (Shore et al. 2016) and RS Oph (Osborne et al. 2011). Such variation is likely to be due to dense blobs of ejecta passing through the line of sight. A sporadic quasi-periodic oscillation of the SSS X-ray flux with a central period of 71 ± 2 s and an amplitude of up to 15 per cent between days 188 to 193 was discovered by Page, Beardmore & Osborne (2015). Similar QPOs have been seen in the SSS X-ray emission from other novae (e.g. Osborne et al. 2011, Ness et al. 2015, Beardmore et al., in preparation); their origin, due to white dwarf spin or atmospheric oscillations, has not been definitively identified.

The temperature of the X-ray emission at the start of the SSS interval was found to be low, kT ~ 20-40 eV, eventually rising to kT ~ 60 eV. The TMAP[3] plane-parallel, static, non-local-thermal equilibrium stellar atmosphere grids, often used to fit nova X-ray spectra, do not extend below kT ~ 38 eV, so could not be used to fit the X-ray spectra of this nova. A blackbody (BB) model was therefore used to fit the white dwarf photosphere spectrum, with the absorbing column NH set to $1.4 \times 10^{21}$ cm$^{-2}$ (Section 3.1). There were sufficient counts at energies above the SSS that an underlying optically thin component (APEC in XSPEC) was also required to fit the full 0.3-10 keV spectra; this likely arises from shocks within the ejecta. Assuming a distance of 1.2 kpc (Section 3.1), the luminosities of both the white dwarf and shock components are shown in panels 3 and 5 of Fig. 4 respectively. The combination of declining BB luminosity and rising temperature, interpreted at face value, points to a photospheric radius shrinking 2-3 orders of magnitude from a highly inflated value down to a more typical white dwarf radius by the end of the observations, consistent with post-outburst expectations.

*2.3. Hubble Space Telescope STIS Observations*

On 2015 November 6 (day 236.3), a medium resolution spectrum was obtained with the *STIS* instrument on the Hubble Space Telescope (Figures 6 and 7). As reported by Kuin et al.

---

[3] TMAP: Tübingen NLTE Model Atmosphere Package: http://astro.unituebingen.de/~rauch/TMAF/flux_HHeCNONeMgSiS_gen.html

(2015b), the spectrum covers the wavelength range 1130-3110 Å, and further details regarding the reduction and analyses of these data will be published by Shore et al. 2018.

### 2.4. Ground-based Observations from Mt. Abu

Near-IR 0.8 – 2.4 µm spectroscopy at a resolution of ~1000 and broad band JHK photometry of V5668 Sgr were carried out on several epochs between 2015 October and 2016 June (see Table 4 and Figure 8) using the 1.2m telescope of the Mount Abu Infrared Observatory (Banerjee & Ashok 2012) with the near-IR camera/spectrograph (NICS) which uses a 1024 × 1024 HgCdTe Hawaii array (Banerjee and Ashok 2012). These data were collected after the abatement of the monsoon at Abu and are in addition to those reported in Banerjee et al (2015). The observational technique and data analysis has been previously described in summarizing our similar observations of Nova V339 Del (Gehrz et al. 2015) and greater details can also be found in Banerjee and Ashok (2012) and Banerjee et al. (2015). The standard stars used for the observations reported here were SAO 187239 (spectral type B8III) for spectroscopy and SAO 186704 (Spectral type A3 III) for photometry. Both these standards have very similar sky coordinates as the nova and were observed at similar air masses as the nova. The Mt Abu data are being analyzed with CLOUDY in a separate work and will be released later.

### 2.5 Ground-based Optical Spectroscopy from MDM and MMT

Optical spectra of V5668 Sgr were obtained on four epochs between 2015 June and September with the 2.4-m Hiltner telescope of the MDM Observatory on Kitt Peak and with the 6.5-m MMT on Mount Hopkins (see Figures 9, 10, and 11). The MDM spectra were obtained on 2015 June 7.391 and September 12.181 UT using OSMOS (Ohio State Multi-Object Spectrograph; Martini et al. 2011). A 1.2″ wide entrance longslit combined with a high— efficiency, low—resolution VPH grism and a 4064 x 4064 pixel STA CCD were employed covering the 3980-6860 Å spectral region at a nominal spectral resolution of 3.8 Å. Spectra of HgNeArXe spectral line lamps and a quartz--halogen lamp were obtained to provide wavelength calibration and allow the removal of pixel-to-pixel variations in the response on the detector. Several spectra of spectrophotometric standard stars (BD+33 2642, BD+25 3941, Hiltner 102, and G191B2B) were obtained during each night to remove the instrumental response function and provide flux calibration.

The MMT spectra were obtained on 2015 June 17.380 and 18.361 UT with the Blue Channel Spectrograph (Schmidt et al. 1989). A 1″ x 180″ longslit was used with a 500 line per mm grating and a thinned STA 2688 513 x512 pixel detector covering all or part of the 3800 to 7100 Å region at a nominal resolution of 3.6 Å. A UV36 longpass filter was used to block 2nd order light from contaminating the red portion of our spectra. As in the MDM spectra, the spectrum of a HeArNe lamp provided wavelength calibration while the spectrum of a quartz-halogen lamp provided flatfield correction images. A spectrum of the spectrophotometric standard star Kopff 27 was obtained during each night to provide flux calibration.

All optical data from both instruments were reduced using standard IRAF packages[4] and standard spectral extraction and calibration techniques.

3. PHYSICAL PARAMETERS ASSOCIATED WITH THE OUTBURST

*3.1 Time of Outburst, Reddening, and Distance*

---

[4] IRAF is distributed by the National Optical Astronomy Observatory, which is operated by the Association of Universities for Research in Astronomy (AURA) under a cooperative agreement with the National Science Foundation.

We take the time of outburst to be 2015 March 15.634 UT (JD 2457097.13; day zero), the maximum light to be $m_B = +4.59$ and $m_V = +4.39$ and the expansion velocity of the slow-moving ejecta to be 530 km s$^{-1}$ (see Banerjee et al. 2016),  If the ejecta have an equatorial disk and bi-polar lobes, there may be low density, high velocity material as well, but there is no evidence of the high velocity material in any of the IR spectra we present here. Most of the dust may be expected to reside in the dense equatorial waist which is moving at the lower velocities. The 530 km s$^{-1}$ velocity is consistent with the line widths in all IR spectra reported in this paper.

The Ly-α line was well exposed in the HST STIS spectrum and used as an independent means to derive the interstellar component of the extinction.  Since V5668 Sgr formed dust, there could be issues with any extinction derived from the 2175Å feature since it might still (at the STIS epoch) be partly contributed by the nova ejecta.  The Leiden/Argentine/Bonn (LAB) all sky 21-cm line survey (Kalberla et al. 2005) was used to obtain integrated neutral hydrogen column density and line profiles toward the nova.  The 21 cm emission, however, is that along the *entire* sight line through the Galaxy.  Therefore, using the weak neutral interstellar absorption lines (e.g. O I 1302Å) and the strong neutral and singly ionized resonance transitions (e.g. O I 1302Å, C II 1334,1335Å), the relevant velocity interval was chosen,  and using the wings of the interstellar medium H Ly-α profile we found a neutral hydrogen column density of  $N_{HI} = 1.4 \times 10^{21}$ cm$^{-2}$. This was used to model the observed Ly-α given the Bohlen et al. (1978) calibration to derive an interstellar extinction to the nova of E(B-V) = 0.3 ± 0.05. The comparison between the predicted, computed Ly-α profile based 21 cm calibration and the STIS spectrum is shown in Figure 12.

Using the visual light decline rate of 100 ± 10 days for V5668 Sgr (Banerjee et al. 2016),  the MMRD relation given in Della Valle & Livio (1995) gives an absolute visual magnitude of $M_v = -6.9 \pm 0.4$. Applying the reddening derived above to the visual maximum, we find the distance to be in the range 0.98 to 1.45 kpc. We assume that $D = 1.2 \pm 0.2$ kpc in the discussion that follows.

### 3.2 Outburst Luminosity

Assuming optically thick spherically symmetric ejecta that completely cover the central engine, the outburst luminosity, $L_o$, is given by:

$$L_o = 4.03 \times 10^{17} D^2 [(\lambda F_\lambda)_{max}] L_\odot \qquad (1).$$

where D is the distance to the nova in kpc, and $(\lambda F_\lambda)_{max}$ is the Planckian emission maximum in W cm$^{-2}$. We determine $(\lambda F_\lambda)_{max}$ by fitting a blackbody curve to the AAVSO data at maximum light (see Figure 13). The fit for B = 4.59, V = 4.32, R = 4.07, and I = 3.87 and a reddening of $E(B-V) = 0.3 \pm 0.05$ from the HST STIS spectrum (Figure 13), gives T = 14,633 ±1830 K, $(\lambda F_\lambda)_{max}$ = 2.1 ± 0.8 x 10$^{-13}$ W cm$^{-2}$, and $L_o = 1.28 \pm 1.06$ x 10$^5$ $L_\odot$ which is about a factor of two above the Eddington limit for a Chandrasekhar mass WD. The derived outburst luminosity for V5668 Sgr is high when compared to the observed outburst luminosities cited by Gehrz (1999) for the dusty CO novae NQ Vul (7 x 10$^4$ $L_\odot$), LW Ser (4 x 10$^4$ $L_\odot$), N1668 Cyg (10$^5$ $L_\odot$), and V705 Cas (5.5 x 10$^4$ $L_\odot$). We assume $L_o$ = 1.28 x 10$^5$ $L_\odot$ in the calculations that follow below.

The soft X-ray flux (0.3-1 keV) of V5668 Sgr is seen to rise by around four orders of magnitude between days 100 to ~200, falling back most of the way over the following year. The harder X-ray flux (2-10 keV) also rises and falls over this interval, but is clearly behaving independently of the softer flux (Fig 4). The soft X-ray flux is characterized by an initially low blackbody temperature, kT ~ 20-40 eV, rising to 50-70 eV as the luminosity starts to decline (Fig 4). This is the photospheric emission from the hot WD, which continues to burn residual fuel after the initial outburst. The properties of this 'super-soft' X-ray emission have been linked (by models) to the mass of the WD and the mass ejected in the eruption (e.g., Schwarz et al. 2011). Wolf et al. (2013, 2014) calculate the duration and effective temperature of the super-soft phase for a range of CO and ONe nova. They find that higher mass WDs burn at a higher temperature, and that all super-soft emission timescales are shorter at higher WD mass. For V5668 Sgr, we find that $L_{max}$ ~ 5.4 x $10^4$ $L_\odot$, $kT_{max}$ = 50-70 eV, $t_{SSS, on}$ = 168 days past outburst, and $t_{SSS,off}$ = 240-340 days past outburst. This luminosity is larger than the luminosity in any model presented by Wolf et al. (2013, 2014), and is likely an over-estimate to the luminosity resulting from blackbody fitting (e.g ., Krautter et al. 1996). The turn-off times and peak temperatures are consistent with WD masses in the range 1.1-1.2 $M_\odot$.

## 4. DISCUSSION

### *4.1 Dust Formation and the Mass of the Dust*

A very deep dust formation event, similar to those seen in NQ Vul (Ney and Hatfield 1978), LW Ser (Gehrz et al 1980a), and V705 Cas (Shore et al, 1994; Mason et al. 1998; Evans et al. 1997, 2005), occurred in V5668 Sgr during 2015 June and July (see Figure 2).  Our first  SOFIA FORCAST observation was executed at JD 2457178.9 (2015 June 6 UT; day 81.7); this was *within hours* of dust nucleation which, from the intersection of linear fits to the immediate pre- and post-dip in the visual light curve, we estimate occurred at JD 2457178.5 (day 81.4 +1.9,-7.7). On day 81.7, the IR SED showed a smooth continuum due to thermal emission from dust with hydrogen recombination lines from the hot gas faintly visible above the dust continuum.   The second SOFIA FORCAST observation was executed at JD 2457209.87 (2015 July 6 UT; day 112.75), within ~10 days of the maximum dust optical depth; we estimate, by fitting a parabola to the bottom of the deep minimum in the light curve (see Figure 1b), that this occurred at JD ~ 2457221.4 (2015 July 17.9 UT; day 114.2). Our K-band Mt Abu/SOFIA IR spectrum on Day 81.7 was well fitted by a blackbody spectrum of T ~1090 K, near the ~1000 K condensation temperature that is consistent with observations of dust formation in other novae (Gehrz 1999, Lynch et al. 2008, Evans and Gehrz 2012). Indeed, as these IR observations seem to have caught V5668 Sgr within hours of dust condensation (see above), this must be a reasonable estimation of the condensation temperature of the dust.

The conditions for the formation of graphitic or amorphous carbon dust in stellar outflows have been discussed by Gail and Sedlmayr (1984). They find that if a carbon atom has time to migrate over the grain surface and finds the energetically most suitable site to bind to the surface, a crystalline grain ensues, otherwise the grain is amorphous. In addition, they show that polycrystalline grains form if the condensate is depleted before the temperature has dropped below ~1100 K, otherwise the grain is amorphous. We note that the carbon dust in the nova V339 Del was carbonaceous , and condensed at 1480 K (Evans et al. 2017), whereas the dust in V5568 Sgr condensed at 1090 K. The featureless continuum and lack of detectable silicate dust emission in excess of the local 10 micron continuum suggests that the dominant grain composition condensing in the expanding ejecta was amorphous carbon material.

The slightly higher condensation temperature (see Figure 2) as compared to that observed in other novae that have formed carbon dust (e.g. NQ Vul, Ney and Hatfield 1976; LW Ser, Gehrz et al. 1980a; V705 Cas, Mason et al. 1998; V2362 Cyg, Lynch et al. 2008) and the deviation of the spectrum below the blackbody continuum at long wavelengths may suggest the presence of small amorphous carbon grains (for which the $\beta$-index is $\approx 0.754$ – see below) or a dust temperature gradient in a shell that is optically thin at long wavelengths. Small grains would be expected in the initial phase of a condensation event.

By Day 112.75 the grain temperature had fallen to ~720 K, the spectrum was well fitted by a blackbody curve out to ~20 μm, the bolometric dust emission had increased by a factor of ~ 3.3 , and the hydrogen recombination lines present in the day 81.7 SOFIA/FORCAST spectrum (see Figure 2) were blanketed by the dust emission continuum. The slight deviation of the SED below the Rayleigh-Jeans tail of the 720 K blackbody beyond wavelengths longer than ~20 μm is consistent with the grains having grown significantly larger than they had been on 2015 June 5. The blackbody dust emission maxima were $(\lambda F_\lambda)_{max}$ = 6.3 x $10^{-15}$ W $cm^{-2}$ and 2.1 x $10^{-14}$ W $cm^{-2}$ on Days 81.7 and 112.75 respectively. The smooth dust continua on both days are consistent with the assumption that the primary grain constituent was carbon.

The Mt Abu spectra taken between 2015 October and 2016 May are shown in Figure 8. The notable feature of the four spectra taken between days 206 and 235 in 2015 October and November is the steep rise of the continuum towards the red in the H and K bands due to thermal emission from dust. We have fitted blackbodies to the continuum in the HK-band data obtained at Mt Abu. While these cover a narrower waveband than the SOFIA data the Mt Abu data provide a reasonable estimate of the dust temperature as they generally fall on the Wien tail of the emission (see Evans et al, 2017). However, the absence of data at longer wavelengths is likely to result in an overestimation of the temperature. Blackbody fits were made to the Mt Abu data are shown in Figure 8. This blackbody fits fail in the J band region due to additional contributions from the central engine WD and bremsstrahlung emission from gas that have not been included here. We find that the four spectra between days 206 and 235 are reasonably well fitted with blackbody temperatures ranging from ~950K to ~1000 K.

The temporal evolution of the emission maximum in $(\lambda F_\lambda)_{max}$ and the dust temperature, as determined from both SOFIA and Mt Abu data, are shown in Figures 14a and 14b and listed in Table 5. Notwithstanding the fact that K-band data alone may overestimate the grain temperature, there is a temperature minimum around day 106 (when we have SOFIA data only). Such a dust temperature minimum has been seen in novae with deep dust minima before, in NQ Vul and LW Ser (see Bode & Evans 1983, Gehrz et al. 1980). The "depth" of the temperature minimum in NQ Vul and LW Ser was ~250 K, as it is in V5668 Sgr. Destruction of the newly-condensed grains is required to account for this phenomenon, (misleadingly called the "isothermal" phase), as rigorously shown by Mitchell et al. (1983). This is further discussed below.

Evans and Gehrz (2012) have shown that the dust mass in a dust shell that is optically thin in the IR is given by:

$$M_{dust} = 4.74 \times 10^{21} \frac{\rho_d D^2 (\lambda F_\lambda)_{max}}{A T^{(\beta+4)}} M_\odot \qquad (2)$$

where T is the dust temperature (in K), D is the distance is in kpc, $\rho_d$ is the dust grain density in gm cm$^{-3}$, and $(\lambda F_\lambda)_{max}$ is the Planckian maximum (assuming blackbody continuum emission) of the dust emission in units of W cm$^{-2}$. *A* and β are defined in such a way that the Planck mean absorption efficiency for the dust is $\langle Q_{abs} \rangle = aAT^\beta$ with the grain radius *a* in cm; we note that the dust mass is independent of *a*. As discussed by Gehrz and Ney (1992; see their Appendix A), $(\lambda F_\lambda)_{max}$ is a well-known proxy for the "bolometric" flux for blackbody emission.. In what follows we assume amorphous carbon dust (see above), for which the density is 1.85 g cm$^{-3}$ (see e.g. Rouleau & Martin 1991). Adopting the Planck means for amorphous carbon tabulated by Blanco et al. (1983) and fitting power-law functions over the temperature range 400 – 1700K, we find Plank mean absorption cross sections of $\langle Q_{abs} \rangle = 58.16aT^{0.754}$ for amorphous carbon ,

where the grain radius, a, is in cm (see Evans et al. 2017). These cross sections, together with equation (2), yield the dust masses for D = 1.2 kpc listed in Table 5 and shown in Figure 14c).

There is a significant increase in dust mass from day 81.7 (condensation) until day 112.75, accompanied by a cooling of the grains; following this there is a warming of the dust grains and a decrease in dust mass. The latter may be due either to a decrease in the number, or size, of the grains, or both, and has previously been noted for the dust forming CO novae NQ Vul (Ney and Hatfield 1978) and LW Ser (Gehrz et al. 1980a). We stress that, despite the rather large systematic uncertainty in the absolute value of the dust mass resulting from the uncertainties in distance and reddening, the photometric errors that establish the warming of the dust and its decrease in mass are small.

The carbon dust mass that condensed in the ejecta of V5668 Sgr at maximum (~$1.2 \times 10^{-7}$ $M_\odot$) was comparable to the amount of carbon dust that condensed in the ejecta of the CO nova V705 Cas (~$2.1 \times 10^{-7}$ $M_\odot$; Shore at al. 1994) and ~ 4 times higher than the amount of carbon dust that condensed in the ejecta of the recent CO nova V339 Del (~$3 \times 10^{-8}$ $M_\odot$, see Evans et al. 2917). Assuming efficient condensation of carbon in ejecta with a gas to dust ratio that is consistent with solar abundance (200:1), the ejected gas mass of V5668 Sgr would have been 200 $M_{dust}$ ~ $2.4 \times 10^{-5}$ $M_\odot$, a typical value for a CO nova (see Gehrz 1999).

*4.2 Grain Size*

The maximum radius to which the grains grew in the ejecta of V5668 Sgr can be estimated from the ratio of the outburst luminosity to the dust shell luminosity at IR maximum, $f = (L_{dust})_{max}/L_o$, assuming that the outburst luminosity was maintained by the central engine for at least several months as in other CO novae (see Gehrz et al. 1998, Gehrz, 1999, Evans & Gehrz 2012). Setting the area covered by the grains equal to the fraction of the luminosity intercepted and reradiated by the dust, we have:

$$f 4\pi R^2 = f 4\pi (V_o t)^2 = N \pi a^2 \qquad (4)$$

where $V_o$ is the outflow velocity of the slow-moving ejecta (cm s$^{-1}$), R is the dust shell radius (cm) at a time t (sec) after the outburst, $N$ is the total number of grains in the dust shell, and $a$ (cm) is the grain radius. Using the form for the Planck mean absorption cross section that we have derived in Section (4.1) above, the dust luminosity is given by:

$$L_{dust} = N 4\pi a^2 <Q_{abs}> \sigma T^4 = N 4\pi a^3 A \sigma T^{(\beta+4)} \qquad (5)$$

where $\sigma = 5.7 \times 10^{-5}$ erg s$^{-1}$ cm$^{-2}$ deg$^{-4}$ is the Stefan-Boltzmann constant. Solving equations (4) and (5) for the grain radius, $a$, yields:

$$a = \frac{L_{dust}}{16\pi f (V_o t)^2 A \sigma T^{(\beta+4)}} = \frac{L_o}{16\pi (V_o t)^2 A \sigma T^{(\beta+4)}} = \frac{L_o}{16\pi R^2 A \sigma T^{(\beta+4)}} \qquad (6)$$

where $L_o$ is the bolometric luminosity of the nova. On day 81.7, we have $R = 3.77 \times 10^{14}$ cm for $V_o = 530$ km s$^{-1}$, $T=1090$ K, and with $L_o = 1.28 \times 10^5$ $L_\odot$ we have $a = 0.78$ µm. As the SOFIA observation was carried out within hours of grain nucleation, grain growth was likely extremely rapid. Shore and Gehrz (2004) have described the physical conditions that might promote rapid grain growth. For day 112.75, we have $R = 5.13 \times 10^{14}$ cm and $T = 720$ K, giving $a = 2.9$ µm, showing clear evidence of further grain growth. However, by day 206 – 235 we find that $a \sim 0.2$ µm, assuming that the nova retains constant bolometric luminosity over this time. The implication is that substantial grain destruction occurred after the dust maximum. If constant bolometric luminosity was not maintained, then using $L_{dust}$ and assuming $f < 1$ yields grain sizes of $a > \sim 0.8$ µm. The evolution of grain size with time is shown in Fig. 14d).

The number of grains that condensed in the ejecta of V5668 Sgr may be estimated from:

$$N_d = \frac{3M_d}{4\pi a^3 \rho} \quad (7)$$

where $M_d$ (gm) and $a$ (cm) are already determined above. The values of $N_d$ are given in Table 5 and its dependence on time shown in Fig. 14e). There is clearly a substantial increase in grain number between condensation and maximum optical depth, after which the number seems to level off. This implies that the decrease in dust mass is primarily due to a decrease in grain size rather than a decrease in grain number. We stress that, despite the rather large systematic uncertainty in the absolute values of the grain size and grain number resulting from the uncertainties in distance and reddening, the photometric errors that establish the temporal trends in these quantities are small.

The decrease in dust mass and grain size must be due to grain destruction. Gehrz et al. (1980a) attributed these effects to sputtering of the grains by hard radiation from the hot WD and the decrease in the vapor pressure in the expanding shell, while Mitchell et al. (1984, 1986) have considered the effect of grain charge on dust destruction. In the case of V5668 Sgr, an additional cause of dust destruction could have been irradiation by the X-ray radiation that was observed during the visual recovery (see Figure 3).

The mechanisms for the destruction of dust by X-ray radiation have been discussed in detail by Fruchter et al. (2001) in the context of dust destruction by gamma-ray bursts and other energetic events. They show that the survival of grains depends on the parameter $E_{51}/D_{100}^2$, where $E_{51}$ is the energy emitted in the form of X-rays at 1 keV, normalized to $10^{51}$ erg, and $D_{100}$ is the distance of the dust from the X-ray source in units of 100 pc. For V5668 Sgr this parameter has the value ~1.0 x $10^6$ if it is assumed that the bulk of the WD emission is in the form of X-rays at the time of maximum dust emission (112.75 days) and that the grains are exposed to X-rays for

~1 month. Figure 1 of Fruchter et al. shows that carbon dust with a radius ~ 1 μm will not survive exposure to X-rays, consistent with our conclusion that there is a period of substantial grain destruction after day 112.75. Our treatment does not allow for the fact that the X-ray emission in V5668 Sgr is likely to be thermal rather than power-law (as in Fruchter et al.'s analysis), that we have estimated the time for which the grains in V5668 Sgr are exposed to X-radiation and that grains with $a > 1$ μm are not considered by Fruchter et al. (see below). However, we do not expect that our conclusion regarding grain survival in the nova environment is likely to be seriously in error.

The temporal evolution of the dust formation event is also demonstrated in Figure 3 where the near IR colors are plotted versus time. Figure 3 uses data up to day 107.32 from Banerjee et al. (2015) and the remaining data from this work as presented in Table 2. For example, an increase can be seen in the (J-K) color which reaches a maximum around July-August, coincident with the deep dip in the optical light curve, and which is then followed by a gradual decrease as the dust is subsequently destroyed.

Dust emission is seen to blanket the emission lines in the H and K bands to a large extent in the 2015 and 2016 NIR spectra. A magnified view of the H and K spectra of 28 February 2016 shows the presence of Brackett series hydrogen recombination lines (Br 10 to Br 16) in the H band and in the K band the presence of the Brγ (2.1656 μm) and HeI 2.0581 μm line. The IJ band (0.8 to 1.4 μm), which is less influenced by dust emission, shows the emission lines more clearly with the HeI 1.0831 μm line being extremely strong with a peak-to-continuum ratio in the range 30 to 50. The strong HeI emission indicates high excitation conditions and suggests that the ejecta were possibly in the nebular stage at this phase. This is supported by the *HST/STIS* spectrum taken on 2015 Nov. 9 that showed He II 1640Å and other high ionization species dominating the ultraviolet emission spectrum.

A detailed identification of lines in the spectrum of 2016 Feb 28 is shown in Figure 15. At this stage, coronal lines began to be prominently seen in the spectrum (Banerjee et al. 2016). We detect [Ti VII] 1.715 µm, [Si VI] 1.9641 µm, [Ca VIII] 2.3205 µm and a strong line at 1.55 µm that has been previously attributed to [Cr XI] 1.5503 µm (Wagner and Depoy, 1996) . The Br 10 line at 1.7362 µm appears stronger than expected, which could be due to a contribution from [P VIII] expected at the same wavelength.  There are 3 unidentified coronal lines at 1.110 µm, 1.190 µm and 2.090 µm. The former two lines have appeared several times earlier in nova spectra (e.g. Venturini et al. 2004 and references therein; Banerjee, Das and Ashok 2009) but have remained unidentified. The 2.090 µm feature was seen earlier in V1974 Cyg  (Wagner and Depoy, 1996),  and V959 Mon (Banerjee, Ashok and  Venkataraman, 2012).  It has tentatively been attributed to [Mn XIV] at 2.092 µm (Wagner and Depoy,1996).   The coronal lines have a distinct profile, best illustrated by the [Si VI] line profile which has two strong blue and red peaks with a very deep dip in between at line center (Figure 16).  The dip almost reaches the continuum level and might have done so in a higher resolution spectrum. This profile shape is in stark contrast to the He and H profiles (Figure 16) which show a central peak at the line core and two shoulders on the blue and red side.  The other coronal lines also share this double peaked structure; a feature which is useful in identifying them unambiguously. Thus the 2.090 µm line, whose true nature and origin had been discussed earlier (e.g. Banerjee, Das and Ashok , 2009) , is now unambiguously established as a coronal line.

The critical density for $10^4$ K gas at which collisional de-excitation starts becoming important is ~ $10^8$ $cm^{-3}$ for the [SiVI] 1.9641 µm line (Greenhouse et al. 1993; Spinoglio 2013). Hence this forbidden line essentially traces relatively low density gas and in this context, the two distinct peaks of the profile are likely associated with low density bipolar lobes. In such a geometry, shaping of the ejecta takes place by a dense equatorial torus which prevents expansion of the ejecta in the equatorial plane while permitting a fast, ballistic expansion into the polar regions to form the dilute and low density bipolar lobes. On the other hand, the dust could either reside in clumps or in the thick equatorial torus, both of which could be dense enough to prevent the UV and soft X-rays penetrating sufficiently deeply to destroy the dust.

*4.3 Dust Formation and the Optical Spectroscopic Signature*

Our MDM spectrum from 2015 June 7 (JD 2457180.5; day 83.8; Figure 9 lower) was obtained just prior to the precipitous (or order 5 magnitudes at V) decline in brightness of the light curve associated with the dust formation event. At this epoch the spectrum is dominated by permitted emission lines of the Balmer series and Fe II (multiplets 37, 38, 42, 48, 49, 55, and 74) with P-Cygni absorption components. In addition, forbidden lines of [O I] 6300 Å and 6363 Å are present and [N II] 5755 Å may also be present indicating the beginning of the evolution into the nebular phase. The profiles of the [O I] lines are relatively symmetric with an average full-width zero intensity (FWZI) of 1880 km s$^{-1}$.

The MMT spectra obtained on 2015 June 17 and 18 (JD ~ 2457191; day 95; Figure 9 middle) are similar to the spectrum obtained at MDM on 2015 June 7 some 10 days earlier. The MMT spectra were obtained during the decline of the visual light curve as a result of the dust formation event, which bottoms out near ~day 108. The forbidden lines of [O I] 6300 Å and 6363 Å have increased in strength relative to the Balmer lines and [N II] 5755 Å was now present.

The spectrum obtained on 2015 September 12 (~day 150; Figure 9 upper) differs significantly from those obtained earlier in 2015 June at the onset of the dust formation event.

The spectrum exhibits strong permitted emission lines of the Balmer series including H$\alpha$, H$\beta$, H$\gamma$, and H$\delta$; He I 4471 Å, 5875 Å, and 6678 Å; He II 4686 Å and weak 5411 Å; N II 5679 Å, 5938 Å, and 6482 Å; N III 4640 Å, and C I 4267 Å. In addition, strong forbidden lines are also present in the spectrum indicating that V5668 Sgr entered the nebular phase sometime before 2015 September 12 but after 2015 June. Strong forbidden lines present in our spectra include [O I] 6300 Å, 6363 Å, and the auroral line at 5577 Å; [N II] 5755 Å , and [O III] 5007 Å, 4959 Å, and 4363 Å. Also present for the first time are weak coronal lines (relative to the other forbidden lines) arising from [Fe VII] 6085 Å, 5721 Å, 5277 Å, and perhaps 5159 Å; [Fe VI] 5677 Å, and [Ca V] 5309 Å. The line profiles of both the permitted and forbidden lines consist of multiple components and all exhibit stronger blue peaks than red peaks. The Balmer lines generally consist of at least 3 components while the [Fe VII] 6085 Å line consists of perhaps only 2 components but it is of lower signal-to-noise ratio. The FWZI of H$\beta$, [O III] 5007 Å, and [Fe VII] 6085 Å are 3300, 2550, and 3160 km s$^{-1}$ respectively. These FWZI measures are broader than those observed in the UV spectra with HST/STIS in 2015 November (see below) but qualitatively consistent with the rapidly dropping emissivity of the expanding envelope and the resulting narrowing of the emission lines.

The H$\alpha$ and [O I] 6300 Å line profiles are shown in Figures 10 and 11. In the 2015 September spectra (Figures 10 and 11 lower) both emission line profiles are relatively symmetric with Halpha exhibiting a P-Cygni profile. Ten days later, on 2015 June 17-18, the H$\alpha$ and [O I] profiles (Figures 10 and 11 middle) are noticeably asymmetric to the red with the blue wing or peak stronger in intensity resulting from the increasing opacity in the envelope. By 2015 September, the H$\alpha$ emission line (Figure 10 upper) exhibits a prominent castellated line profile with at least 5 distinct components. The [O I] line profile (Figure 11 upper) exhibits a smoother line profile but some evidence for multiple emission components.

The HST (+STIS) spectra were obtained after the visual light curve recovered from the deep optical minimum and was on a plateau from ~day 160 through 210. The STIS emission line profiles typically show two peaks with the blue peak brighter than the red peak. The full line width of the Mg II 2800 Å, C II 2324 Å, and N II 2143 Å lines is $2000 \pm 100$ km s$^{-1}$. We observe lines from multiple ionization stages: CII, CIII, CIV; NII, NIII, NIV, and NV; OI, OII, OIII, and OV, which evidently means that the ejecta consist of matter at a wide range of ionizations, quite possibly this is due to clumping in the ejecta. Similarly, the ground-based optical line profiles provide evidence for clumpy expanding ejecta. However, the shapes of the line profiles are significantly affected by the presence of dust along the line of slight.

The asymmetry in the Si II 1526 line profiles in the STIS spectrum, like the He II 1640 Å profile, are best explained by a bipolar ejecta .

The H Ly-α line is comprised of a saturated absorption line with two emission peaks offset by about the same velocities as the peaks in the other emission lines (Figure 7). These peaks are consistent with the 530 km s$^{-1}$ expansion velocity derived from the IR spectra, and indicate that the bulk of the ejected mass is associated with this velocity. The line profiles observed in the STIS spectrum are evidence for a bipolar geometry of the ejecta. Yet the dust is changing the ionization state in those ejecta, which implies that enough dust is formed between the bipolar ejecta and the WD to cause this effect. In the light curves of both continuum and line flux we see occasional excursions. We searched for a characteristic time scale but were unsuccessful.

*4.4 Dust Formation and the UV Evolution*

The dust dip centered on day 110 is followed by a UV flux rise to day 170 of a factor ~40. Over the same interval, the soft X-ray flux rose by a factor of ~300 (see Fig 4). The soft X-ray flux continued to rise to day ~220, while the UV plateaus after day 170. The UV flux rise corresponds to 4 mag. And, as the dust column density is proportional to the extinction in magnitudes, if this rise is the result of a reduction in extinction, the dust column should have decreased by a factor of 4 also. The V-band brightness increase over the same interval was 3.2 mag, significantly larger than would have been expected given the UV rise and the extinction ratio A(291nm)/A(V) ~ 2 valid for both Milky Way and LMC extinction. Dust grains below ~1 micron are optically thin to X-rays (e.g. Fruchter et al. 2001), the atoms trapped in such dust absorb X-rays to the same extent whether they are in dust grains or not. Dust destruction of typical grain sizes back-lit by X-rays likely is not accompanied by a change of X-ray flux. The rise in the soft X-ray flux, by virtue of the physics and the observed properties, is not due to simple dust destruction. In fact the soft X-ray rise from day ~140 is interrupted by deep short-lived dips. This suggests the presence of significant clumping in the ejecta passing through the line of sight. Indeed, the widely accepted explanation for the eventual rise of the soft X-ray flux is the lowering gas column density as the ejecta expand combined with the higher temperatures that are thereby revealed. At least, between days 170 - 240, the soft X-ray emission, as modelled by a blackbody, doubles in temperature, easily accounting for the observed soft X-ray flux rise in this interval. If the dust grains were large enough to be optically thick to X-rays, then the material hidden from the X-ray flux in the grains would have been released on their destruction. This extra material would cause increased X-ray absorption, the precise opposite of what is seen. Dust destruction is expected from the X-ray fluence, and is responsible for the optical-UV flux increase after day 110, but the X-ray flux change is unrelated to this, being due to the hotter and deeper layers of the WD being seen through the declining fog of the thinning ejecta.

Examination of Figure 5 shows that during days 108-129 when the intensity was lowest, the N III flux was lower than the N II flux, while for the rest of the time it was higher. The effect of dust grains larger than the typical wavelengths in the spectrum is to lower the intensity evenly over the spectrum, so it is likely that the change in the ratio of the NII/NIII fluxes was due to a change in the ionization of N. The dust must shield enough radiation from the WD to change the photoionization during the dip. The large intensities of NII and NIII are only possible if the lines are formed in the ejecta.

## 5. CONCLUSIONS

We present 5-28 µm SOFIA FORCAST spectroscopy documenting the early dust formation phase of development of nova V5668 Sgr. Amorphous carbon dust condensation had just commenced 81.7 days after outburst at a temperature of ~ 1090 K, and maximum grain growth was reached around day 112.75 when the shell visual optical depth was $\tau_v$ ~ 5.4 and the shell temperature was 720 K. NIR spectra from Mt Abu, taken between 205.9 to 234.9 days after outburst during the recovery phase show that dust temperature had risen to ~ 950K and that the dust mass had decreased, suggesting that grain destruction occurred during the recovery phase; the prime cause of the decrease in dust mass was a decrease in grain radius rather than in grain number. Dust warming and grain destruction have been observed in several previous dusty CO novae with deep dust minima. A rich emission line spectrum is seen in the 0.8 to 1.4 µm region. Assuming a distance to the nova of 1.2 kpc, we conclude that the grains were amorphous carbon, that the maximum mass of dust produced was 1.2 x $10^{-7}$ $M_\odot$, and that the average grain radius grew to a maximum of a ~ 2.9 µm. The Swift X-ray count rate due to the hot white dwarf peaked around day 220, when its spectrum was a kT = 35 eV blackbody, although it later rose to 50–70 eV. The temperature, together with the super-soft X-ray turn-on and turn-off times, suggests a WD mass ~ 1.1 $M_\odot$. This is the first case of a nova dust event well observed in the UV and X-rays. Our data show that the post-dust event X-ray brightening is not due to dust destruction, which certainly occurred, as the dust is optically thin to X-rays.

# 6. ACKNOWLEDGEMENTS

We thank SOFIA Target of Opportunity Team members Tiina Liimets and Luke Keller for their participation in planning the program and an anonymous referee for helpful comments that substantially improved the presentation. RDG, CEW, and LAH were supported by a USRA SOFIA Cycle 3 Target of Opportunity Nova contract. RDG was supported in part by the United States Air Force and HST GO Program 13828. CEW acknowledges support from HST GO Programs 13388 and 13828. The research at the Physical Research Laboratory is supported by the Department of Space, Government of India. NSF and NASA grants to ASU supported SS. NPMK, KLP, and JPO acknowledge support from the UK Space Agency.

Facilities: SOFIA, Mt. Abu, MMT, MDM, Swift, HST, AAVSO

**Table 1: SOFIA Observation Summary**

| Spectral Element | Config | Inst. | Channel | $\lambda_{eff}$ (μm) | $\Delta\lambda$ (μm) | Coverage (μm) | Slit[a] | R ($\lambda/\Delta\lambda$) | Flight |
|---|---|---|---|---|---|---|---|---|---|
| F111 | Imaging | FORCAST | SWC | 11.1 | 0.95 | … | … | … | F216; F226 |
| G063 | Spect | FORCAST | SWC | … | … | 4.9-8.0 | LS47 | 120 | F216; F226 |
| G111 | Spect | FORCAST | SWC | … | … | 8.4-13.7 | LS47 | 130 | F216; F226 |
| G227 | Spect | FORCAST | LWC | … | … | 17.6-27.7 | LS47 | 110 | F216; F226 |
| G329 | Spect | FORCAST | LWC | … | … | 28.7-37.1 | LS47 | 160 | F216 |
| F315 | Imaging | FORCAST | LWC | 31.5 | 5.7 | … | … | ... | F226 |
| F336 | Imaging | FORCAST | LWC | 33.6 | 1.9 | … | … | … | F226 |
| F348 | Imaging | FORCAST | LWC | 34.8 | 3.8 | … | … | … | F226 |
| F371 | Imaging | FORCAST | LWC | 37.1 | 3.3 | … | … | … | F226 |
| B1_LM | Spect | FLITECAM | … | … | … | 3.3-4.1 | SS20 | 1200 | F244 |
| C2_LM | Spect | FLITECAM | … | … | … | 2.8-3.4 | SS20 | 1300 | F244 |

[a]The LS47 slit is 4.7''x191''

**Table 2: SOFIA Observation Details**

| Flight | Date (UT) | Day Number ($t_0$ = JD 2457103.174) | Spectral Element | Altitude (ft) | ZA (°) | Coadds | Total Int. Time (s) |
|---|---|---|---|---|---|---|---|
| F216 | 2015 June 5 | 75.7 | F111 | 43000 | 62.8 | 5 | 39 |
| | | | G063 | 43000 | 62.9 | 6 | 104 |
| | | | G111 | 43000 | 62.6 | 7 | 297 |
| | | | G227 | 43000 | 62.6 | 9 | 384 |
| | | | G329 | 43000 | 62.7 | 13 | 588 |
| F226 | 2015 July 6 | 106.7 | F111 | 37000 | 39.2 | 4 | 28 |
| | | | F315 | 37000 | 41.8 | 2 | 150 |
| | | | F336 | 37000 | 39.7 | 2 | 153 |
| | | | F348 | 37000 | 40.4 | 2 | 157 |
| | | | F371 | 37000 | 41.5 | 2 | 140 |
| | | | G063 | 37000 | 37.8 | 5 | 151 |
| | | | G111 | 37000 | 36.7 | 8 | 360 |
| | | | G227 | 37000 | 34.5 | 36 | 1195 |
| F244 | 2015 Oct 1 | 193.5 | B1_LM | 40000 | 63.9 | 4 | 240 |
| | | | C2_LM | 40000 | 63.8 | 4 | 180 |

**Table 3: Swift UVOT Continuum Region Fluxes and f1750 and f2143 Fluxes**[5]

| DAY | f204 | f223 | f258 | f291 | f1750 | sig1750 | f2143 | sig2143 |
|---|---|---|---|---|---|---|---|---|
| 92.446 | 1.55E-12 | 8.91E-13 | 1.32E-13 | 1.74E-13 | ... | ... | ... | ... |
| 92.593 | 1.44E-12 | 5.66E-12 | 1.48E-13 | 2.28E-13 | ... | ... | ... | ... |
| 92.646 | 2.57E-12 | 5.43E-12 | 1.98E-13 | 6.41E-14 | 3.24E-10 | 8.78 | 8.43E-11 | 20.12 |
| 93.989 | 3.53E-12 | 2.20E-12 | 1.24E-12 | 5.34E-13 | 1.85E-10 | 3.55 | 2.39E-10 | 18.95 |
| 95.32 | 2.55E-12 | 1.84E-12 | 1.64E-12 | 6.10E-13 | 4.50E-10 | 6.02 | 1.49E-10 | 17.34 |
| 97.309 | 1.49E-12 | 1.31E-12 | 1.78E-12 | 6.89E-13 | 2.35E-10 | 6.23 | 7.42E-11 | 14.09 |
| 97.723 | 1.09E-12 | 1.28E-12 | 1.33E-12 | 4.78E-13 | 1.39E-10 | 6.21 | 7.81E-11 | 21.3 |
| 97.779 | 1.34E-12 | 1.22E-12 | 1.92E-12 | 8.09E-13 | ... | ... | ... | ... |
| 99.109 | 9.91E-13 | 9.01E-13 | 1.37E-12 | 6.42E-13 | 4.09E-11 | 4.25 | 4.21E-11 | 14.96 |
| 99.175 | 9.74E-13 | 8.78E-13 | 1.67E-12 | 7.06E-13 | ... | ... | ... | ... |
| 101.1 | 6.21E-13 | 6.93E-13 | 1.10E-12 | 3.77E-13 | 4.37E-11 | 5.96 | 2.97E-11 | 16.49 |
| 101.245 | 5.05E-13 | 5.42E-13 | 1.07E-12 | 3.33E-13 | ... | ... | ... | ... |
| 103.24 | 1.94E-13 | 1.87E-13 | 4.88E-13 | 1.77E-13 | 2.49E-11 | 10.28 | 9.12E-12 | 12.14 |
| 104.371 | 2.09E-13 | 2.55E-13 | 4.79E-13 | 1.90E-13 | ... | ... | ... | ... |
| 106.422 | 1.64E-13 | 3.12E-13 | 2.61E-13 | 9.85E-14 | ... | ... | 9.26E-12 | 15 |
| 110.078 | 1.19E-14 | 1.77E-14 | 9.67E-14 | 3.75E-14 | ... | ... | 4.53E-12 | 8.01 |
| 110.145 | 1.72E-14 | 2.01E-14 | 1.04E-13 | 3.66E-14 | ... | ... | ... | ... |
| 112.011 | 1.42E-13 | 5.18E-14 | 6.21E-14 | 5.89E-14 | 2.09E-12 | 4.9 | ... | ... |
| 112.073 | 1.09E-13 | 5.43E-14 | 5.75E-14 | 6.25E-14 | ... | ... | ... | ... |
| 112.152 | 1.30E-13 | 5.15E-14 | 6.32E-14 | 6.88E-14 | ... | ... | ... | ... |
| 114.013 | 5.56E-14 | 3.38E-14 | 7.27E-14 | 2.90E-14 | 3.00E-12 | 4.48 | 1.56E-11 | 13.6 |
| 114.081 | 6.29E-14 | 4.24E-14 | 7.51E-14 | 3.16E-14 | ... | ... | ... | ... |
| 114.145 | 6.29E-14 | 4.85E-14 | 9.14E-14 | 3.35E-14 | ... | ... | ... | ... |

---

[5]The continuum band fluxes (erg cm$^{-2}$ s$^{-1}$ Å$^{-1}$) in columns 2-5 are associated with the following bandwidths: 2030-2050 Å, 2220-2240 Å, 2570-2590 Å, and 2900-2920 Å. The integrated line fluxes (erg cm$^{-2}$ s$^{-1}$) and Gaussian widths (Å) given in columns 6 and 8 are for the lines [N III]1750 Å and [N II]2143 Å. The standard deviations (in Å) of the best fits to the lines are given in columns 7 and 9. The full-width-half-maximum (FWHM) of the lines is 2.354 sigma in Å.

**Table 3 (Continued): Swift UVOT Continuum Region Fluxes and f1750 and f2143 Fluxes**

| DAY | f204 | f223 | f258 | f291 | f1750 | sig1750 | f2143 | sig2143 |
|---|---|---|---|---|---|---|---|---|
| 116.132 | 1.39E-13 | 1.00E-20 | 4.07E-14 | 2.39E-14 | ... | ... | 5.73E-12 | 7.97 |
| 116.204 | 8.93E-14 | 4.48E-16 | 4.00E-14 | 2.29E-14 | ... | ... | ... | ... |
| 118.742 | 4.42E-14 | 2.98E-14 | 4.67E-14 | 3.52E-14 | 1.21E-11 | 5.59 | 7.66E-12 | 9.9 |
| 126.712 | 4.25E-14 | 1.41E-14 | 4.74E-14 | 4.38E-14 | 3.41E-12 | 5.07 | 8.85E-12 | 8.12 |
| 126.771 | 5.21E-14 | 2.13E-14 | 4.61E-14 | 4.50E-14 | ... | ... | ... | ... |
| 132.56 | 9.20E-14 | 2.91E-14 | 7.45E-14 | 6.56E-14 | 3.84E-11 | 7.89 | 7.53E-12 | 9.13 |
| 132.623 | 9.75E-14 | 2.86E-14 | 7.37E-14 | 6.42E-14 | ... | ... | ... | ... |
| 140.343 | 8.39E-14 | 1.86E-14 | 5.31E-14 | 1.04E-13 | 6.04E-11 | 6.9 | 1.88E-11 | 8.02 |
| 140.416 | 8.72E-14 | 2.42E-14 | 5.84E-14 | 1.14E-13 | ... | ... | ... | ... |
| 146.403 | 4.90E-13 | 1.39E-13 | 1.24E-13 | 3.92E-13 | 1.49E-10 | 7.3 | 3.01E-11 | 9.78 |
| 146.467 | 4.45E-13 | 1.18E-13 | 1.07E-13 | 3.58E-13 | ... | ... | ... | ... |
| 154.045 | 6.86E-14 | 3.07E-14 | 7.91E-14 | 1.80E-13 | ... | ... | ... | ... |
| 155.509 | 6.82E-14 | 3.34E-14 | 9.92E-14 | 2.03E-13 | ... | ... | ... | ... |
| 160.896 | 4.59E-13 | 1.77E-13 | 2.19E-13 | 5.85E-13 | 3.24E-10 | 5.82 | 4.80E-11 | 8.57 |
| 160.959 | 4.57E-13 | 1.68E-13 | 2.29E-13 | 5.85E-13 | ... | ... | ... | ... |
| 161.229 | 1.07E-13 | 4.87E-14 | 9.49E-14 | 2.41E-13 | 2.35E-10 | 5.3 | 2.94E-11 | 7.5 |
| 167.941 | 5.94E-13 | 3.06E-13 | 5.13E-13 | 7.69E-13 | ... | ... | ... | ... |
| 168.008 | 6.15E-13 | 3.23E-13 | 5.41E-13 | 8.59E-13 | ... | ... | 7.11E-11 | 8.53 |
| 171.732 | 1.95E-12 | 9.61E-13 | 1.37E-12 | 1.04E-12 | ... | ... | 1.51E-10 | 10.63 |
| 174.792 | 8.06E-13 | 4.61E-13 | 6.74E-13 | 8.64E-13 | ... | ... | 8.64E-11 | 9.84 |
| 177.662 | 8.75E-13 | 5.03E-13 | 7.49E-13 | 8.29E-13 | ... | ... | 1.10E-10 | 9.27 |
| 180.848 | 8.75E-13 | 5.26E-13 | 7.16E-13 | 9.14E-13 | ... | ... | 1.06E-10 | 8.31 |
| 184.844 | 4.95E-13 | 3.09E-13 | 5.04E-13 | 7.27E-13 | ... | ... | 8.34E-11 | 8.12 |
| 188.627 | 7.82E-13 | 4.91E-13 | 7.08E-13 | 8.00E-13 | ... | ... | 1.15E-10 | 8.83 |
| 192.289 | 8.43E-13 | 5.28E-13 | 7.08E-13 | 7.60E-13 | ... | ... | 1.18E-10 | 8.61 |
| 196.209 | 9.06E-13 | 5.73E-13 | 7.53E-13 | 7.30E-13 | ... | ... | 1.30E-10 | 8.34 |

**Table 3 (Continued): Swift UVOT Continuum Region Fluxes and f1750 and f2143 Fluxes**

| DAY | f204 | f223 | f258 | f291 | f1750 | sig1750 | f2143 | sig2143 |
|---|---|---|---|---|---|---|---|---|
| 209.043 | 8.03E-13 | 5.13E-13 | 7.32E-13 | 7.96E-13 | ... | ... | 1.27E-10 | 8.02 |
| 223.404 | 9.03E-13 | 5.02E-13 | 8.66E-13 | 1.00E-12 | ... | ... | 1.28E-10 | 7.85 |
| 227.725 | 6.65E-13 | 4.16E-13 | 7.51E-13 | 1.05E-12 | 6.32E-10 | 4.85 | 9.85E-11 | 8.15 |
| 233.377 | 2.25E-13 | 1.96E-13 | 3.18E-13 | 8.46E-13 | ... | ... | ... | ... |
| 236.303 | 7.33E-13 | 4.19E-13 | 7.69E-13 | 9.93E-13 | 1.48E-09 | 5.82 | 1.00E-10 | 8.2 |
| 243.084 | 6.25E-13 | 4.29E-13 | 6.59E-13 | 7.44E-13 | 8.20E-10 | 4.26 | 9.42E-11 | 8.38 |
| 345.738 | 2.25E-13 | 1.96E-13 | 3.18E-13 | 8.46E-13 | 4.80E-11 | 5.7 | 1.62E-11 | 11.13 |
| 345.806 | 2.13E-13 | 1.96E-13 | 2.46E-13 | 7.80E-13 | ... | ... | ... | ... |
| 362.424 | 2.68E-13 | 1.62E-13 | 2.19E-13 | 3.37E-13 | 8.60E-11 | 11.76 | 5.86E-11 | 12.87 |
| 362.486 | 3.07E-13 | 1.50E-13 | 2.27E-13 | 2.38E-13 | ... | ... | ... | ... |
| 385.886 | 2.40E-13 | 1.28E-13 | 2.12E-13 | 2.24E-13 | ... | ... | ... | ... |
| 386.085 | 1.74E-13 | 1.47E-13 | 2.28E-13 | 2.75E-13 | ... | ... | ... | ... |
| 393.79 | 2.50E-13 | 1.72E-13 | 2.51E-13 | 2.92E-13 | 5.98E-11 | 6.17 | 1.31E-11 | 11.78 |
| 393.851 | 2.59E-13 | 1.79E-13 | 2.48E-13 | 2.57E-13 | ... | ... | ... | ... |

**Table 4: Mt. Abu JHK Photometry**

| DATE (UT) | JD | Day[*] | J (error) | H (error) | K (error) |
|---|---|---|---|---|---|
| 2015 Oct 7.56 | 2457303.06 | 205.93 | +8.05 (0.04) | +5.72 (0.05) | +3.65 (0.05) |
| 2015 Oct 18.59 | 2457314.09 | 216.96 | +7.61 (0.06) | +5.90 (0.04) | +4.02 (0.05) |
| 2015 Oct 25.6 | 2457321.10 | 223.97 | +7.70 (0.06) | +6.05 (0.04) | +4.00 (0.10) |
| 2015 Nov 5.55 | 2457332.05 | 234.92 | +7.87 (0.12) | +6.20 (0.13) | +4.17 (0.18) |
| 2016 Feb 28.96 | 2457447.46 | 350.33 | +9.88 (0.08) | +8.84 (0.08) | +7.08 (0.09) |
| 2016 Apr 17.95 | 2457496.45 | 399.32 | +10.30 (0.08) | +9.65 (0.06) | +8.14 (0.07) |
| 2016 Apr 27.91 | 2457506.41 | 409.28 | +10.42 (0.06) | +9.82 (0.05) | +8.33 (0.07) |
| 2016 May 5.9 | 2457514.40 | 417.26 | +10.47 (0.04) | +9.86 (0.05) | +8.45 (0.08) |
| 2016 May 17.88 | 2457526.38 | 429.24 | +10.53 (0.05) | +10.03 (0.06) | +8.70 (0.06) |
| 2016 May 18.92 | 2457527.42 | 430.28 | +10.54 (0.03) | +10.03 (0.04) | +8.74 (0.05) |
| 2016 May 28.91 | 2457537.40 | 440.27 | +10.74 (0.07) | +10.24 (0.03) | +8.82 (0.07) |
| 2016 June 7.03 | 2457546.53 | 449.39 | +10.79 (0.02) | +10.37 (0.08) | +8.97 (0.04) |
| 2016 June 8.02 | 2457547.52 | 450.38 | +10.79 (0.04) | +10.40 (0.04) | +9.10 (0.07) |

[*]Days past outburst

Table 5: Evolution of the dust[*]

| Days from outburst | $T_d$ (K) (± 1σ) | $(\lambda f_\lambda)_{max}$ ($10^{-15}$ W cm$^{-2}$) (± 1σ) | a (μm) (± 1σ) | $M_d$ ($10^{-8} M_\odot$) (± 1σ) | $N_d$ (± 1σ) |
|---|---|---|---|---|---|
| 81.7 | 1090 (3) | 6.31 (0.12) | 0.78 (0.23) | 0.5 (0.16) | 2.75 x $10^{36}$ (0.81) |
| 86.28 | 1038 (15) | 13.3 (0.26) | 0.88 (0.26) | 1.32 (0.63) | 5.03 x $10^{36}$ (1.50) |
| 112.8 | 720 (3) | 21.1 (0.42) | 2.92 (0.86) | 11.9 (4,0) | 1.23 x $10^{36}$ (0.35) |
| 205.96 | 950 (4) | 6.4 (0.13) | 0.23 (0.07) | 0.97 (0.33) | 1.93 x $10^{38}$ (0.57) |
| 216.96 | 957 (17) | 5.0 (0.10) | 0.20 (0.06) | 0.73 (0.37) | 2.20 x $10^{38}$ (0.66) |
| 223.97 | 1005 (8) | 4.0 (0.08) | 0.15 (0.05) | 0.47 (0.18) | 3.43 x $10^{38}$ (1.02) |
| 234.92 | 974 (3) | 3.9 (0.08) | 0.16 (0.05) | 0.53 (0.17) | 3.33 x $10^{38}$ (1.00) |
| 350.33 | 889 (3) | .36 (0.01) | 0.11 (0.03) | 0.08 (0.02) | 1.41 x $10^{38}$ (0.42) |
| 399.32 | 971 (16) | .1 (0.01) | 0.06 (0.02) | 0.014 (0.007) | 1.98 x $10^{38}$ (0.59) |

[*]Errors shown are formal 1 sigma errors from curve and model fitting given the observational uncertainties; values of a, $M_d$, and $N_d$ are for the assumed distance of 1.2 kpc.

FIGURE CAPTIONS

Figure 1a: V-band and UV light curve for V5668 Sgr, data taken from the AAVSO archive (V, black squares) and Swift UVOT (uvw2, red squares). Observation dates for SOFIA (red), Mt Abu (black), MDM (blue), MMT (cyan), HST STIS (blue) and Swift (red) are indicated; for Swift, broken line indicates where the nova was not detected in X-rays. The first SOFIA observation was obtained within hours of dust condensation, the second within 10 days of the deep minimum. The Swift observations were taken at varying cadence over the period indicated, and V5668 Sgr entered the Swift Sun constraint at ~day 240. See text for further details.

Figure 1b: Parabolic fit to the dust formation event showing that maximum dust optical depth in the visible was reached on JD ~ 2457221.4 (2015 July 17.9 UT; day 114.2).

Figure 2: The 5 – 27 μm SOFIA FORCAST spectrum obtained on 2015 June 5 UT (JD 2,457,178.84; day 81.7) during the early condensation phase, showing hydrogen recombination lines from the hot gas superimposed upon the dust continuum.; the SOFIA data are supplemented by a ground-based K-band spectrum, obtained contemporaneously at Mt Abu. The principal quantum series responsible for the IR hydrogen lines present are indicated; note that some are blends. The dust composition appears to be amorphous carbon and the condensation temperature was 1090 K. The 5 – 27 μm SOFIA FORCAST spectrum obtained on 2015 July 6 UT (JD 2,457,209.88; day 112.75) within 10 days of the deep minimum shows that the hydrogen emission lines from the hot gas were blanketed by the dust continuum; the dust luminosity had increased by a factor of ~ 3.3 since June 6. The dust composition continued to be amorphous carbon, and the grains had cooled to 721 K. The lower red curve is the optically thin 5-27 μm

dust continuum from a FORCAST spectrum obtained a year after maximum grain growth on 2016 July 16 UT (JD 2457585.5; day 488.4). At this late epoch, a strong 10 μm silicate emission feature was present, and the luminosity of the dust shell had declined by a factor of ~ 115 from the maximum value. The spectral region strongly affected by telluric ozone is indicated; data in this region have been excised.

Figure 3: Top panel: XRT light curve extracted over the soft (0.3-1 keV) and hard (2-10 keV) bands. The majority of the increase in flux occurred below 1 keV, as the SSS switched on. Panel 2: UVOT continuum flux measurements taken from 20A bins centered on the wavelengths stated obtained from the UV grism spectra (see Table 3). Panel 3: Integrated UVOT line flux measurements for [NIII] 1750Å and [NII] 2143Å (see Table 3) . Panel 4: Optical data taken from the AAVSO database (see also Fig 1). Bottom panel: The near-infrared colors. The J-K color is large (~5.3 mag) during the deep dip in the optical/UV, and then gradually decreases as the IR emission from the dust subsides. Data points prior to day 110 are from Banerjee et al. (2015); the remaining points are from this work.

Figure 4: XRT spectral fits. The fit was that of an optically thin "APEC" thermal component together with a BB component (to model the soft emission) when required. The absorbing column was fixed at $1.4 \times 10^{21}$ cm$^{-2}$. Top panel: 0.3-10 keV XRT light-curve. Second panel: BB temperature. Third panel: Bolometric luminosity of the BB component, assuming a distance of 1.2 kpc. Upper limits were derived assuming BB kT = 20eV. Fourth panel: Temperature of the optically-thin thermal component, modelling the non-SSS emission. Fifth panel: Bolometric luminosity of the optically-thin component.

Figure 5: Swift UVOT grism spectra during the dust formation phase up to the time of the brightness minimum. The spectra near the minimum include data from both grisms, the rest were taken in the UV grism. For some of the spectra, second order lines are contaminated by 294 nm in first order (the 175.0 nm second order line), and 335.0 nm in first order (the 190.9 nm second order line). Parts of the spectra that were affected by zeroth orders from field stars or were too bright for a coincidence-loss correction have been removed.

Figure 6: The HST STIS spectrum covering the wavelength range 1130-3110 Å.

Figure 7: STIS line profiles. The H Ly-α line (bottom panel) is comprised of a saturated absorption line with two emission peaks offset at about the same velocities as the peaks in the He II 1640Å line (top panel).

Figure 8: A series of near-infrared spectra obtained from Mt. Abu between 2015 October and 2016 May are shown. One of the notable features is the strong red-ward rise of the continuum in the H and K band spectra during 2015 October and November. This IR excess, due to thermal emission from dust, is seen to gradually subside with time as the dust grains are destroyed. A blackbody fit at T = 950K is shown in the top panel.

Figure 9: Optical spectra of V5668 Sgr obtained at the MDM (2015 June 7.39 UT and September 12.18 UT) and the MMT (2015 June 17.38 UT and 18.36 UT) Observatories. The June 7.39 UT (day 83.75) spectrum is similar to those of other classical novae early in the outburst. It is dominated by Balmer and Fe II emission lines exhibiting P Cygni-type line profiles. Forbidden [O I] 6300 Å and 6363 Å emission lines are also present. By mid-June, the MMT spectra showed a weakening of the absorption components seen earlier. By September, V5668 Sgr was entering the nebular phase and exhibited emission lines of the Balmer series of H, He I, [N II], and [O III] coinciding with the decline of emission from various Fe II multiplets.

Figure 10: H$\alpha$ line profiles observed during the early evolution of V5668 Sgr and extracted from the spectra shown in Figure 9. The time since the start of the outburst is shown above each line profile. The line profiles through day 94 exhibit P Cygni-type profiles but also show the effects of increasing dust opacity by the attenuation of the line redward of line center and the apparent shift of the line centroid to bluer wavelengths. The profile obtained on day 180 shows multiple resolved components but also the continued presence of a dust.

Figure 11: [O I] 6300 Å line profiles observed during the early evolution of V5668 Sgr and extracted from the spectra shown in Figure 9. The profile observed on day 83 is relatively symmetrical. By days 93 and 94, the profile has become significantly double-peaked compared to H$\alpha$ with some substructure. By day 180, the line is more symmetrical but the centroid is displaced to shorter wavelengths perhaps indicative of continued dust opacity.

Figure 12: Fit of Ly-$\alpha$ line for V5668 Sgr from STIS observation using the column density obtained for the interstellar line intervals and the LAB 21 cm profile (see text). The solid line is the STIS spectrum corrected for E(B-V) = 0.3 extinction which is the same used for the absorption line profile (dashed line).

Figure 13: Blackbody fits to the pseudo-photosphere, using AAVSO data. Red points are observed values; blackbody curve is 8,670 K. Black points are photometry dereddened by E(B − V) = 0.3; solid blackbody curve is 14 633 K and the broken curves are fits to data dereddened by the extreme values of E(B − V) allowed by the uncertainties.

Fig. 14a: Dependence of the maximum infrared emission from the circumstellar dust as a function of time. The errors are small compared to the size of the plotting symbols.

Fig. 14b: Dependence of dust temperature on time. Dust temperature has been determined as described in text. The time of maximum optical depth in the visual light curve is indicated. The error bars reflect the photometric uncertainties.

Fig. 14c: Dependence of dust mass on time. The time of maximum optical depth in the visual light curve is indicated. The error bars reflect uncertainties that are independent of distance and are therefore indicative of the accuracy of the trends exhibited by the plotted parameters.

Fig 14d: Evolution of grain radius with time. The time of maximum optical depth is indicated. The error bars reflect uncertainties that are independent of distance and are therefore indicative of the accuracy of the trends exhibited by the plotted parameters.

Fig 14e: Evolution of grain number with time. The time of maximum optical depth is indicated. The error bars reflect uncertainties that are independent of distance and are therefore indicative

of the accuracy of the trends exhibited by the plotted parameters.

Figure 15: An identification of the lines seen in the NIR spectrum of 2016 February 28 taken from Mt. Abu.

Figure 16: Detail of several line profiles from the NIR spectrum of 2016 February 28 (see Figure 15). There is a considerable difference in the profiles of the hydrogen and helium lines and the coronal [Si VI] 1.9461 μm line.

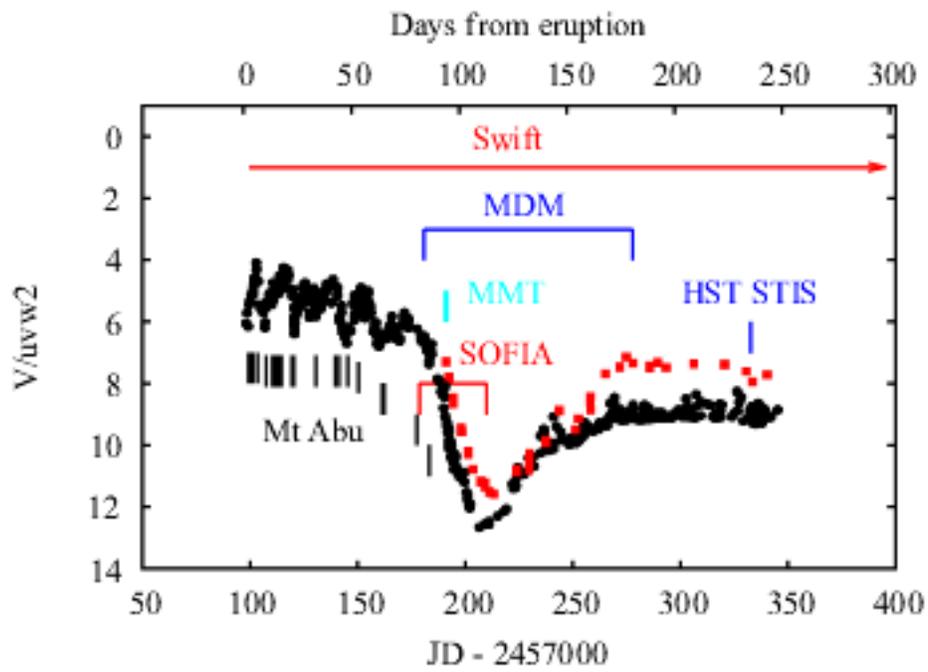

**Figure 1a**

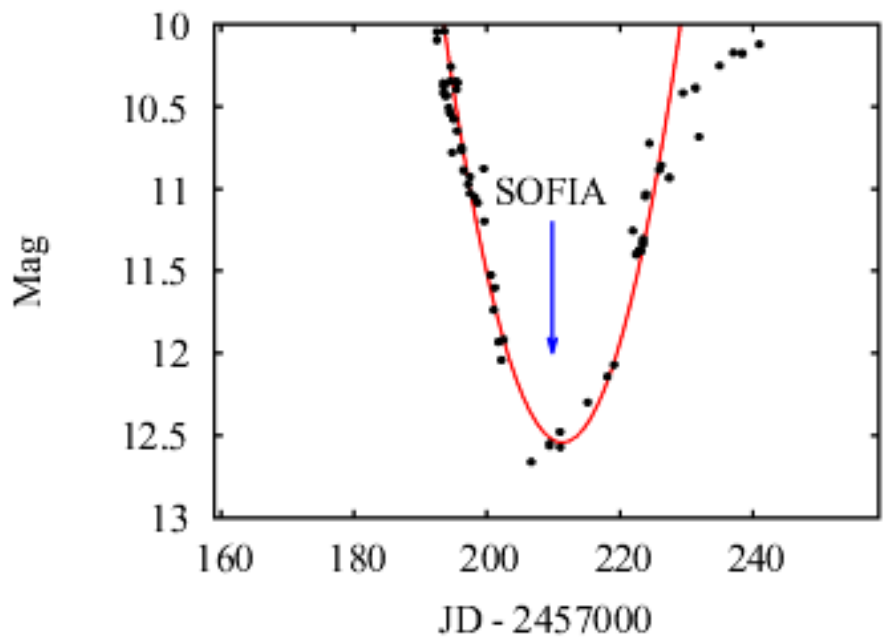

**Figure 1b**

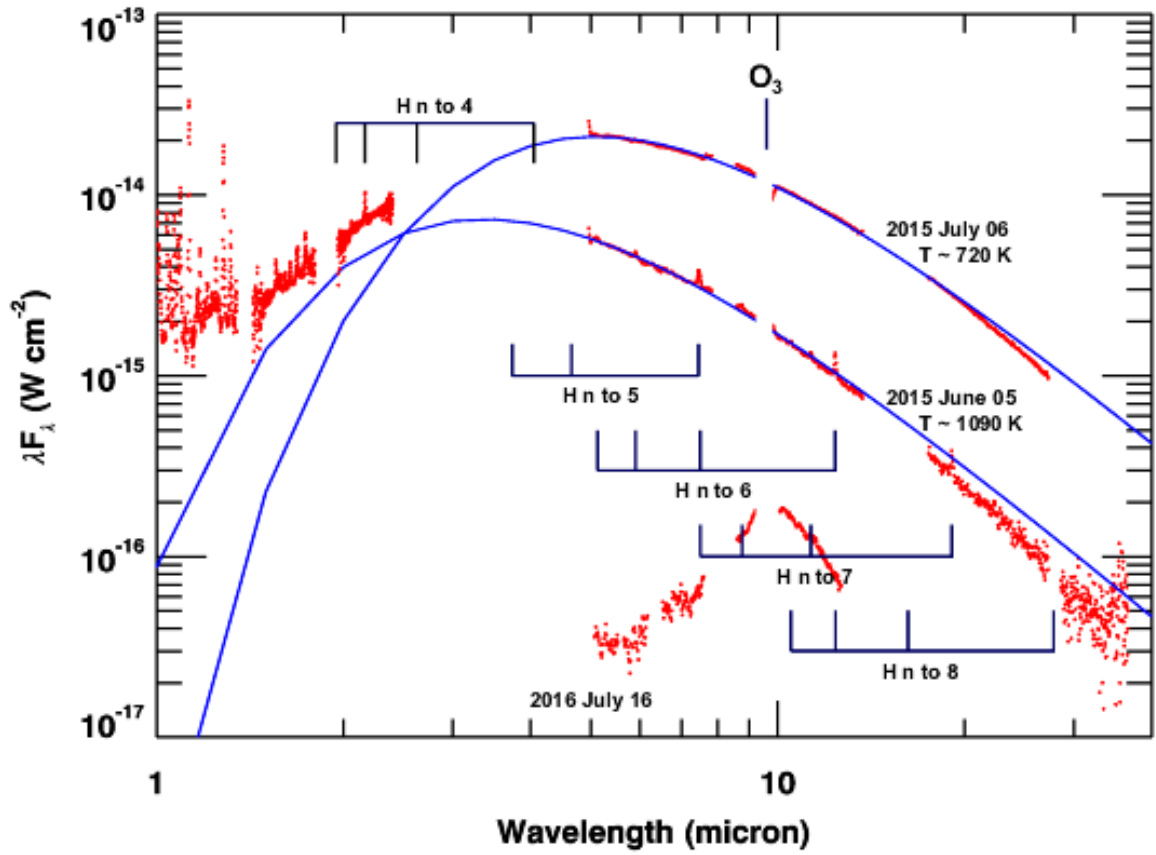

**Fig 2**

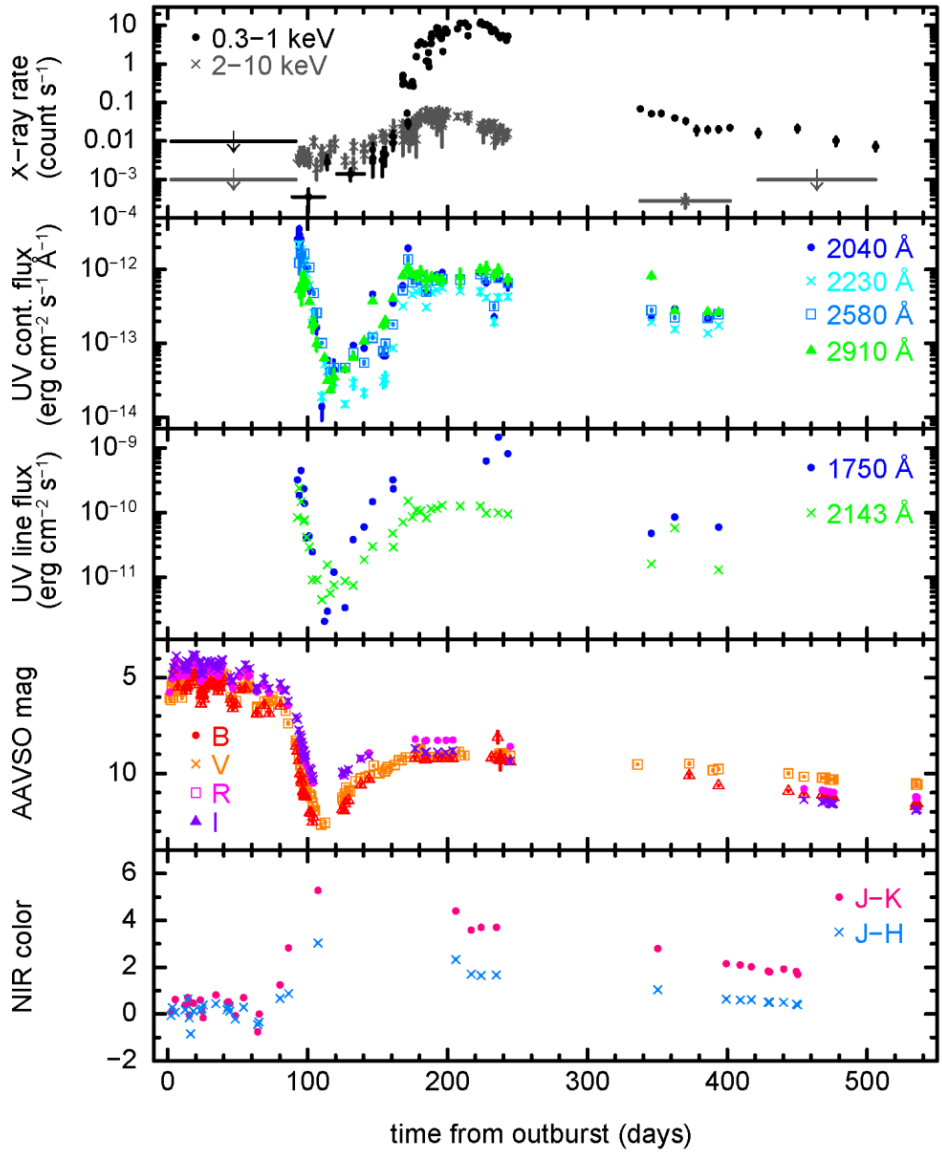

**Fig 3**

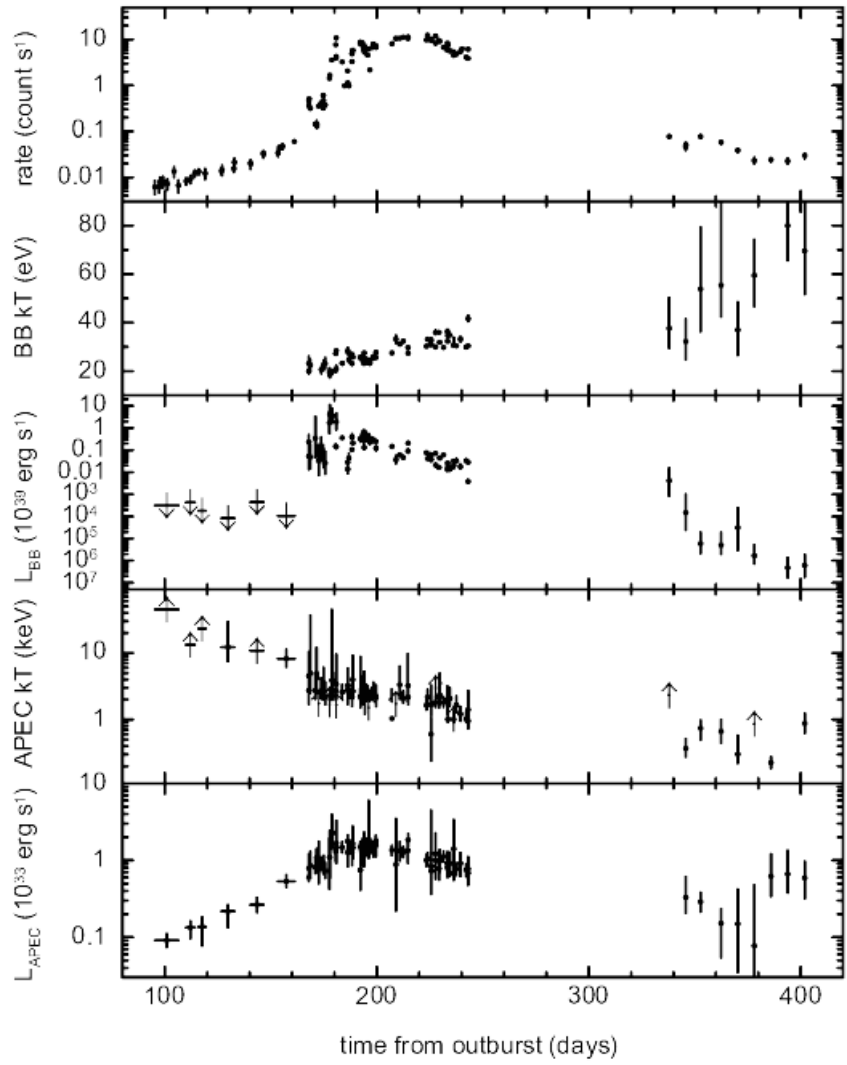

**Fig 4**

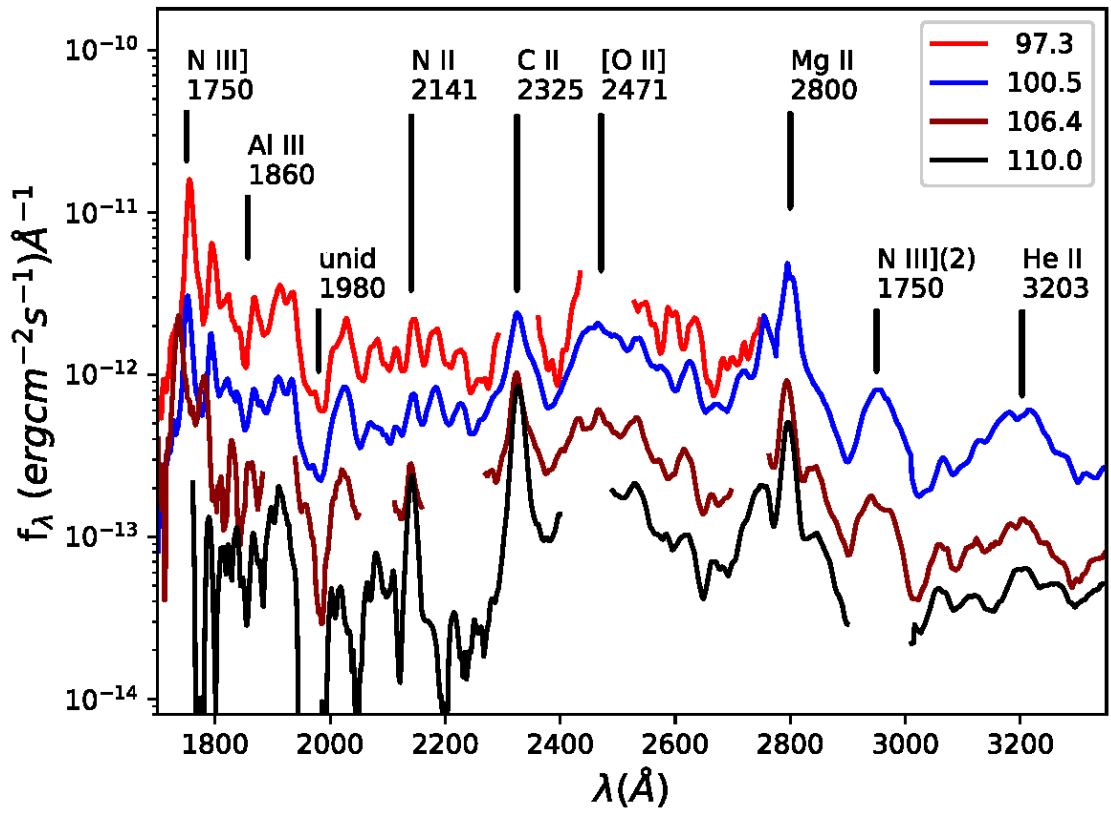

**Fig 5**

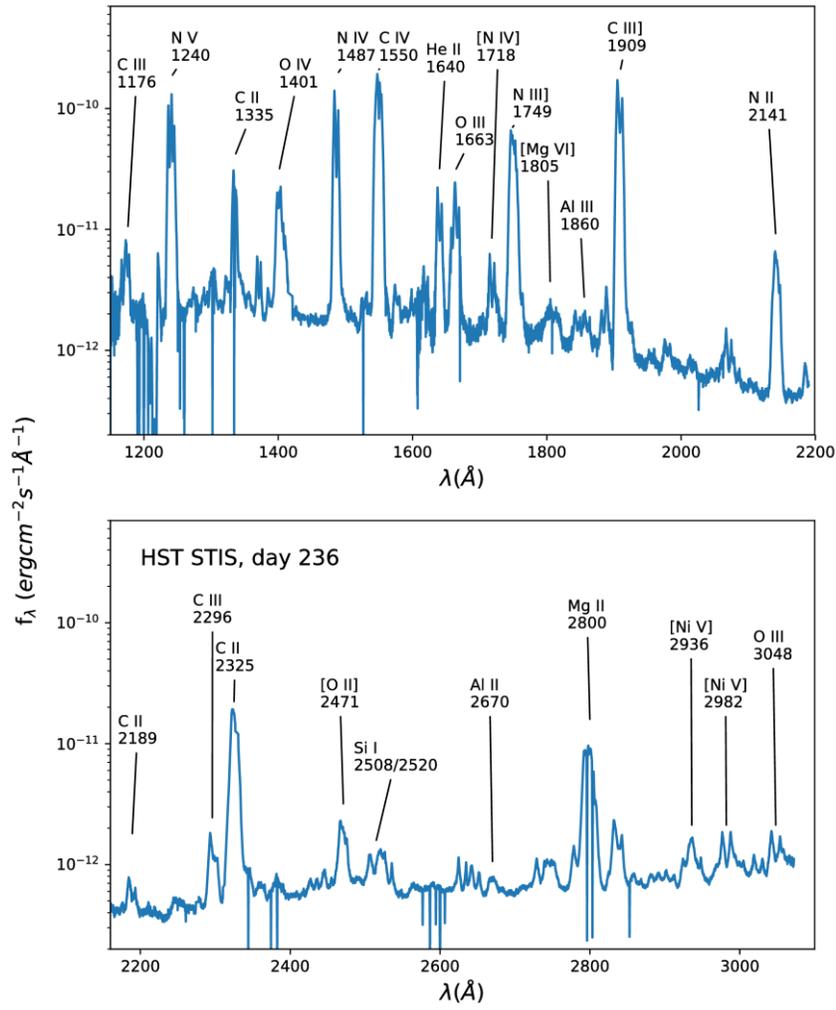

**Fig 6**

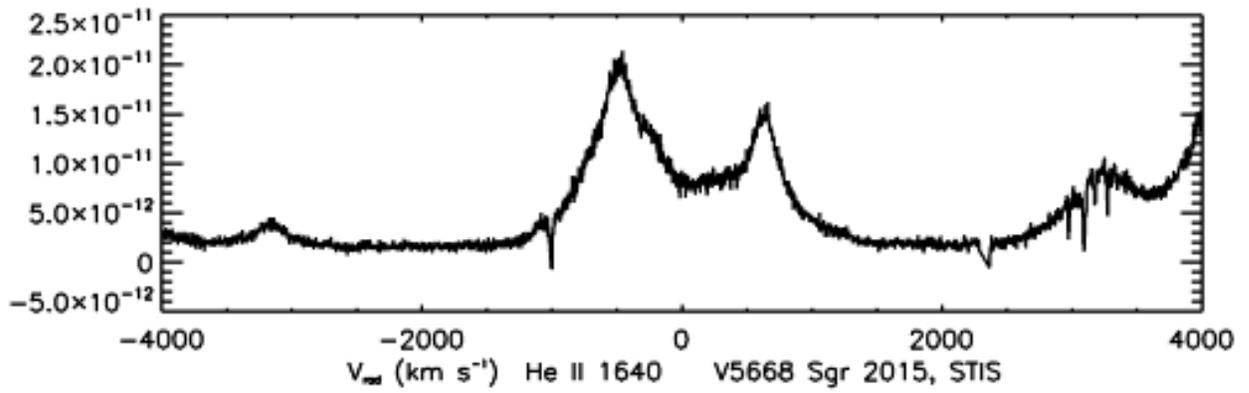
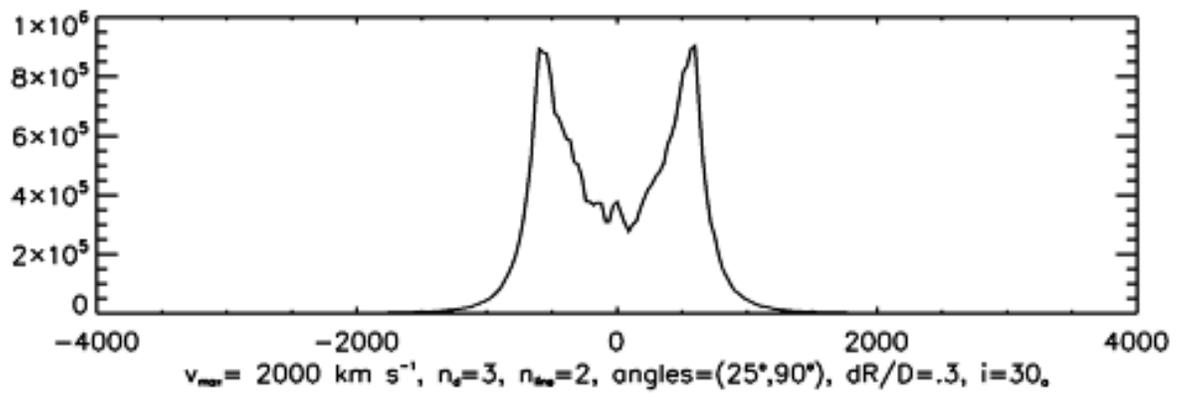

**Fig 7**

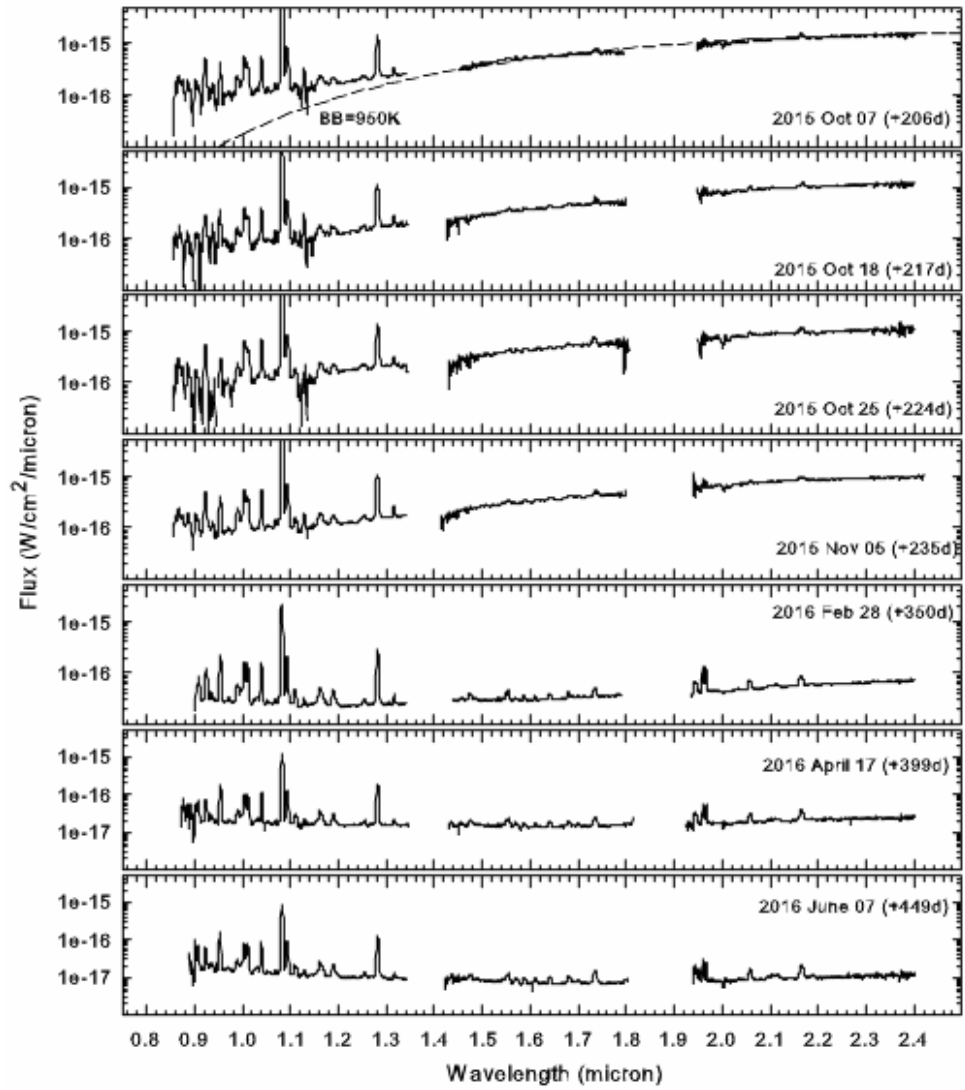

**Figure 8**

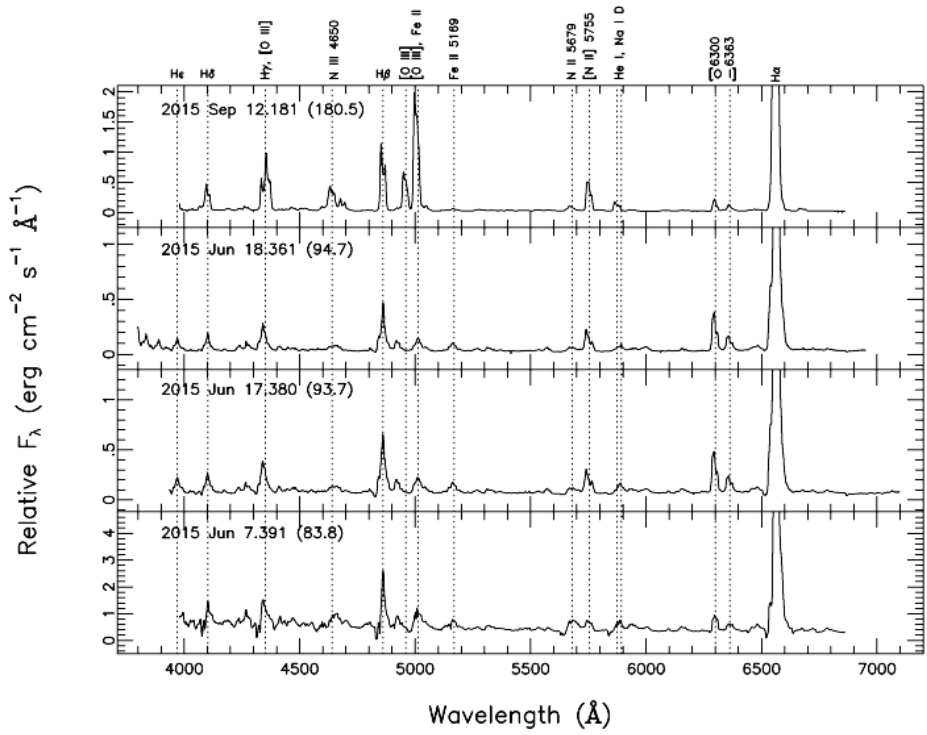

**Fig 9**

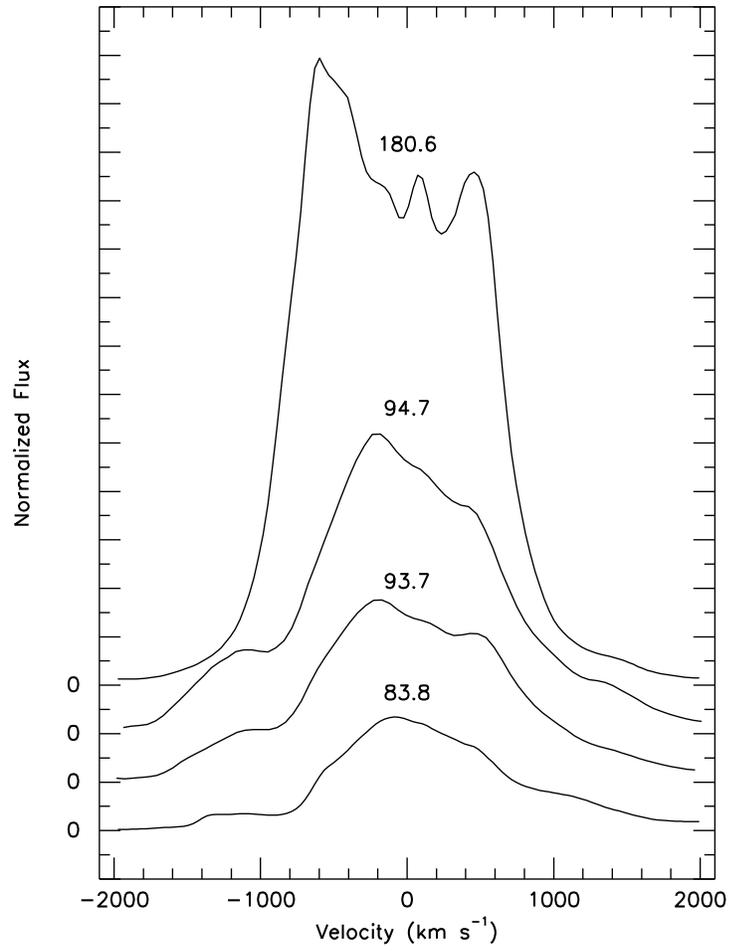

**Fig 10**

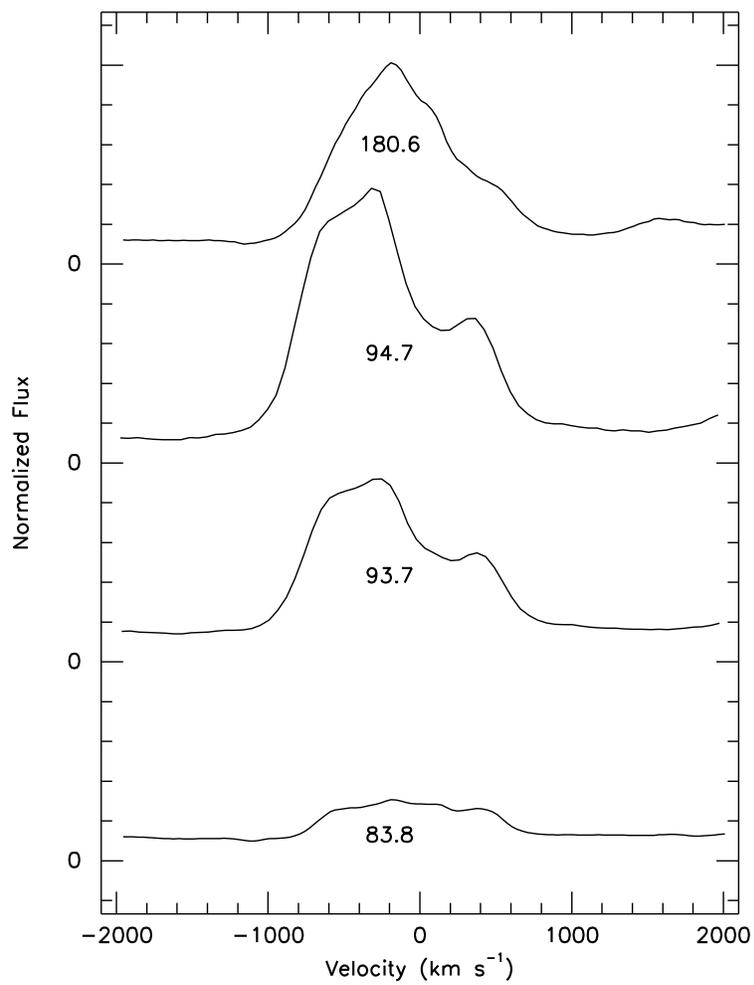

**Fig 11**

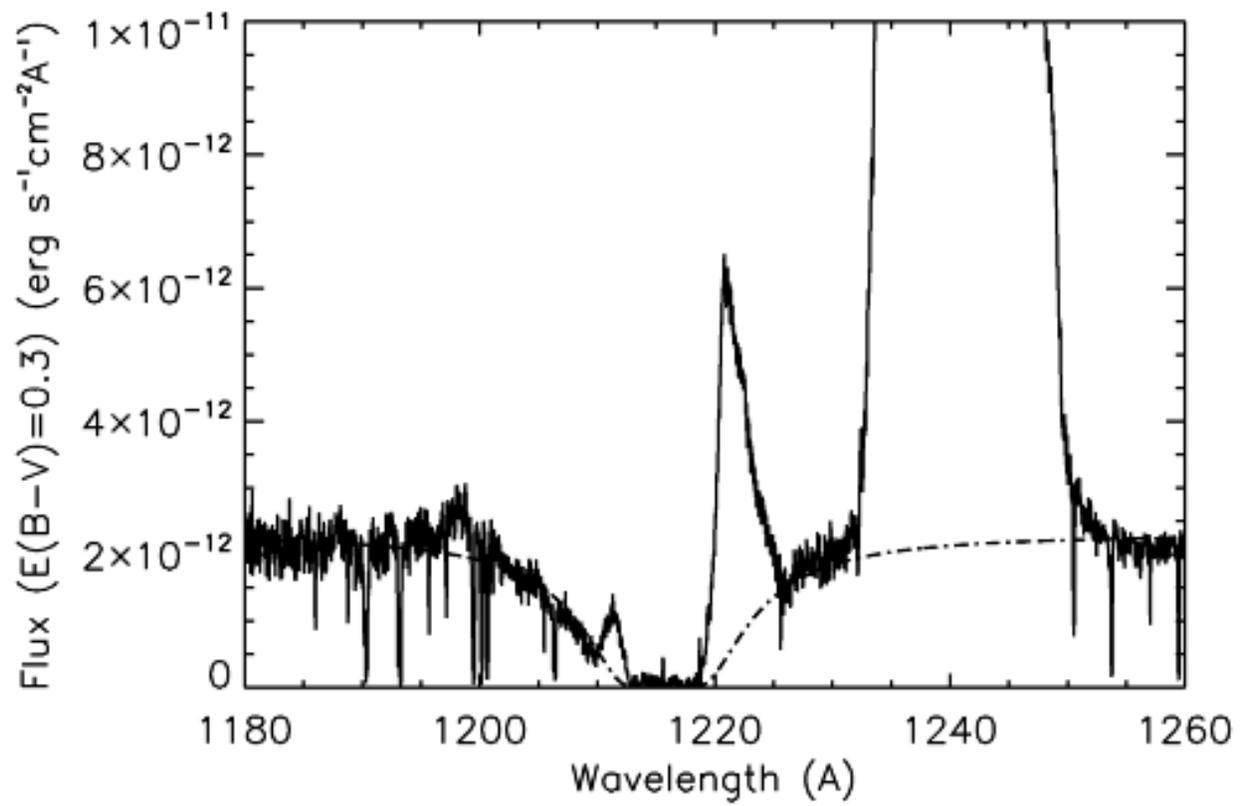

**Figure 12**

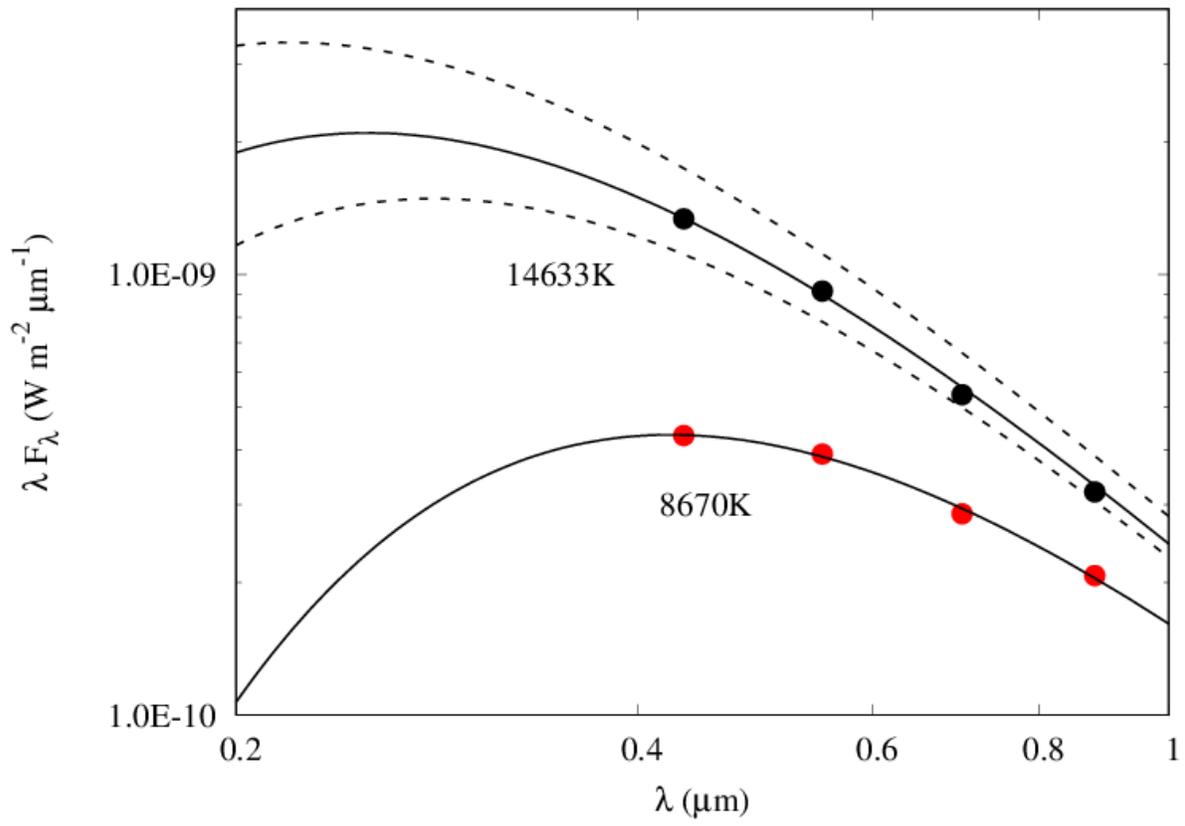

**Fig 13**

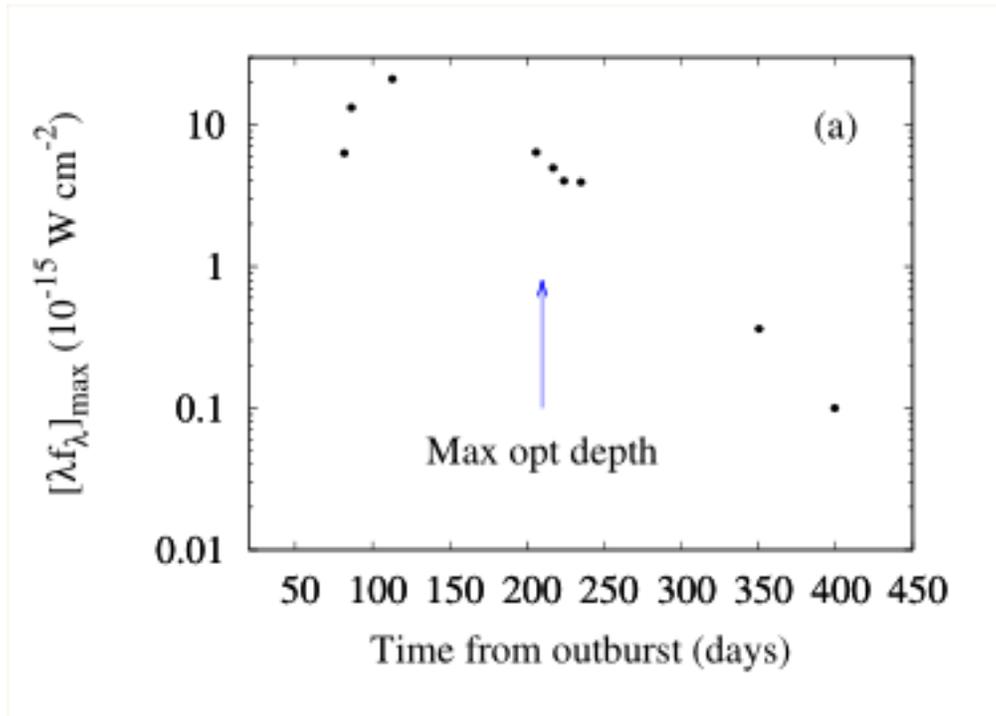

**Fig 14a**

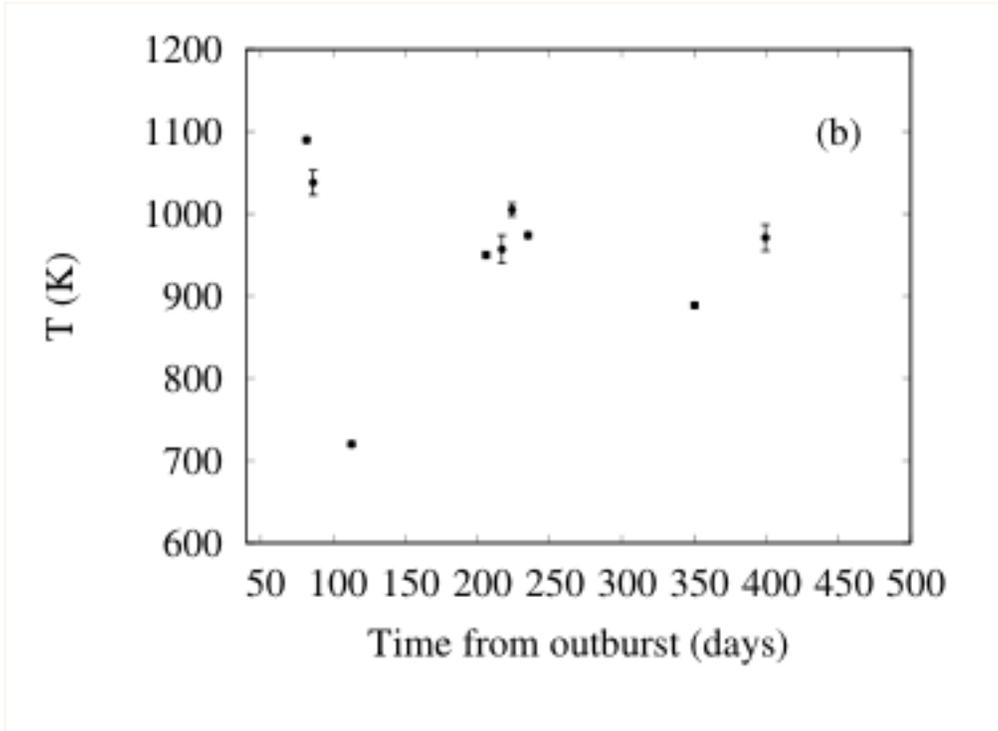

**Figure 14b**

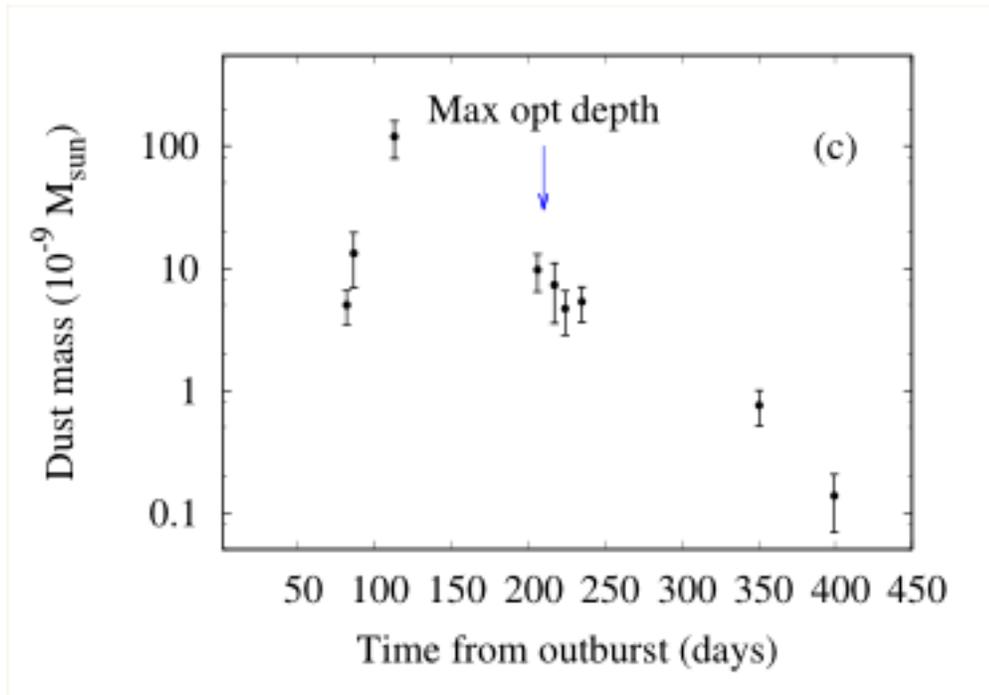

**Figure 14c**

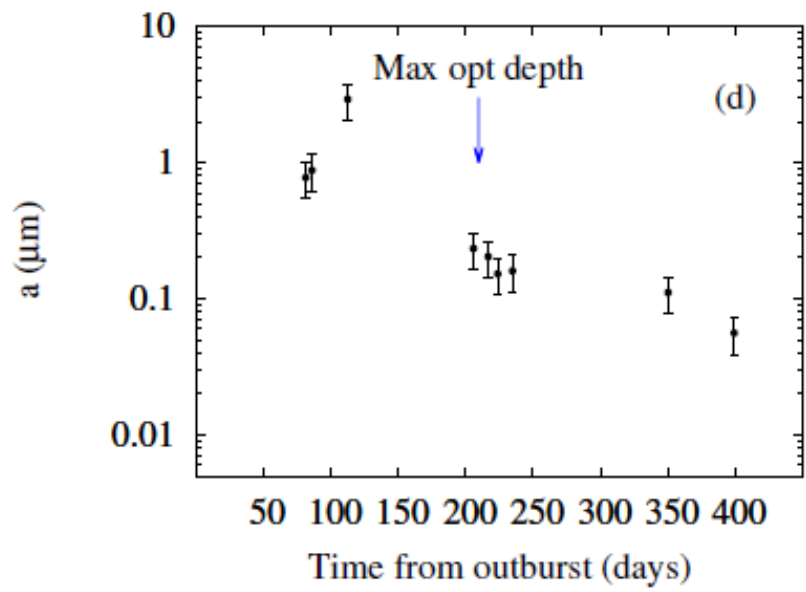

**Figure 14d**

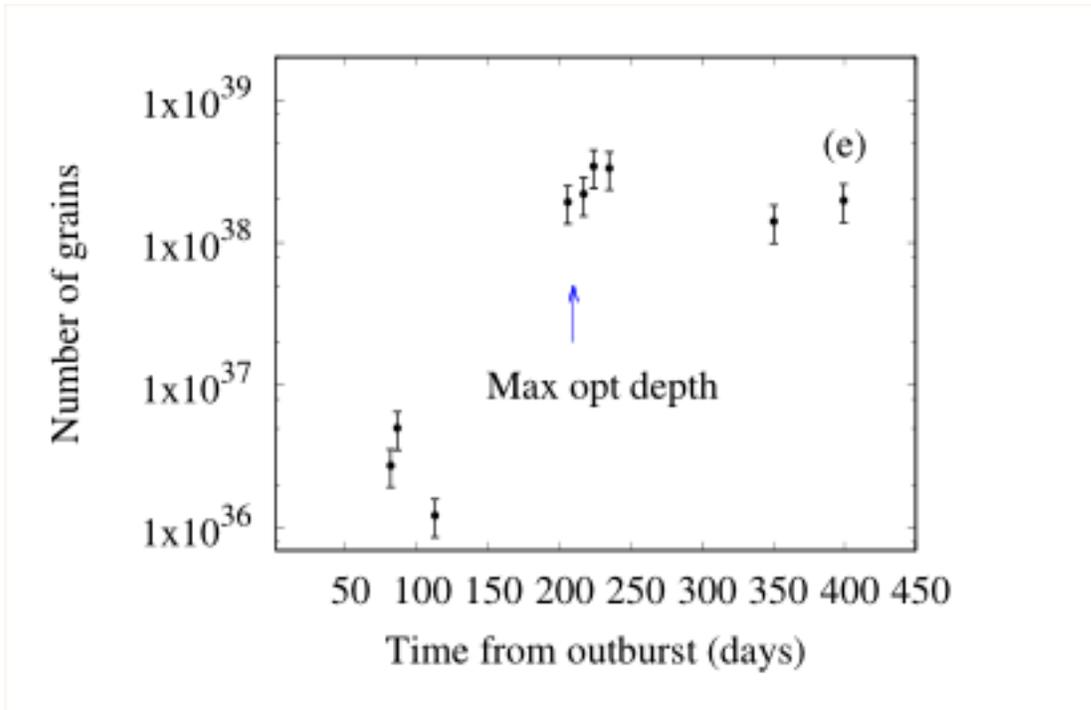

**Figure 14e**

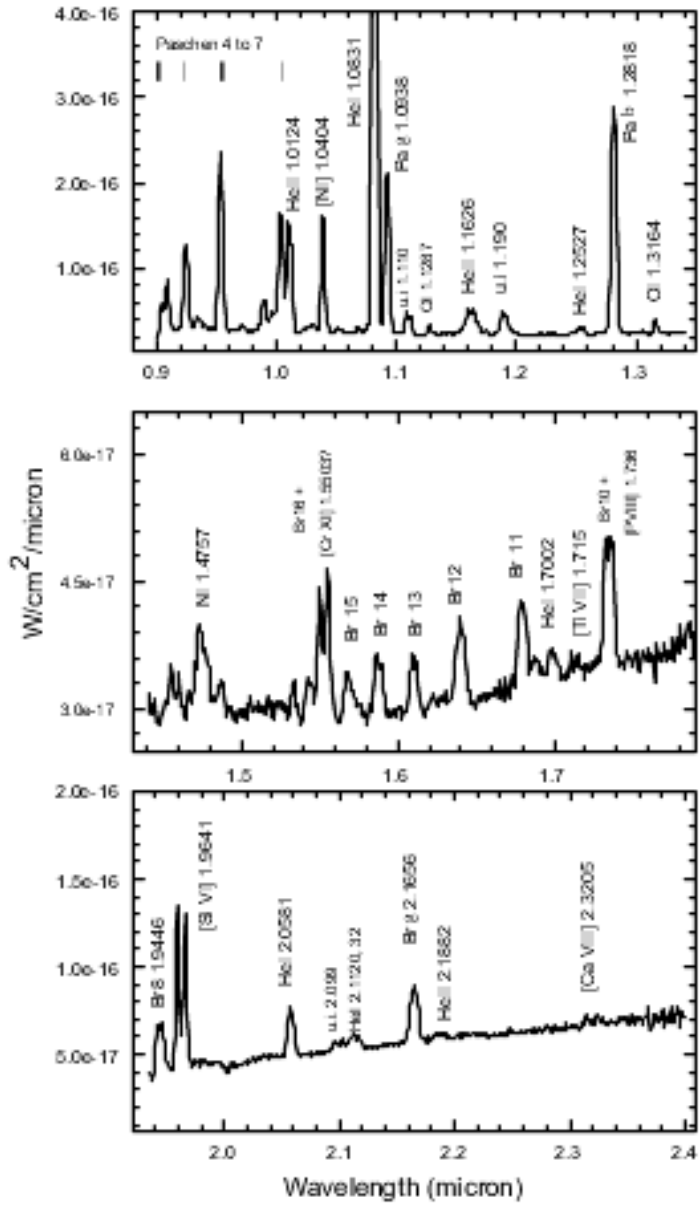

**Figure 15**

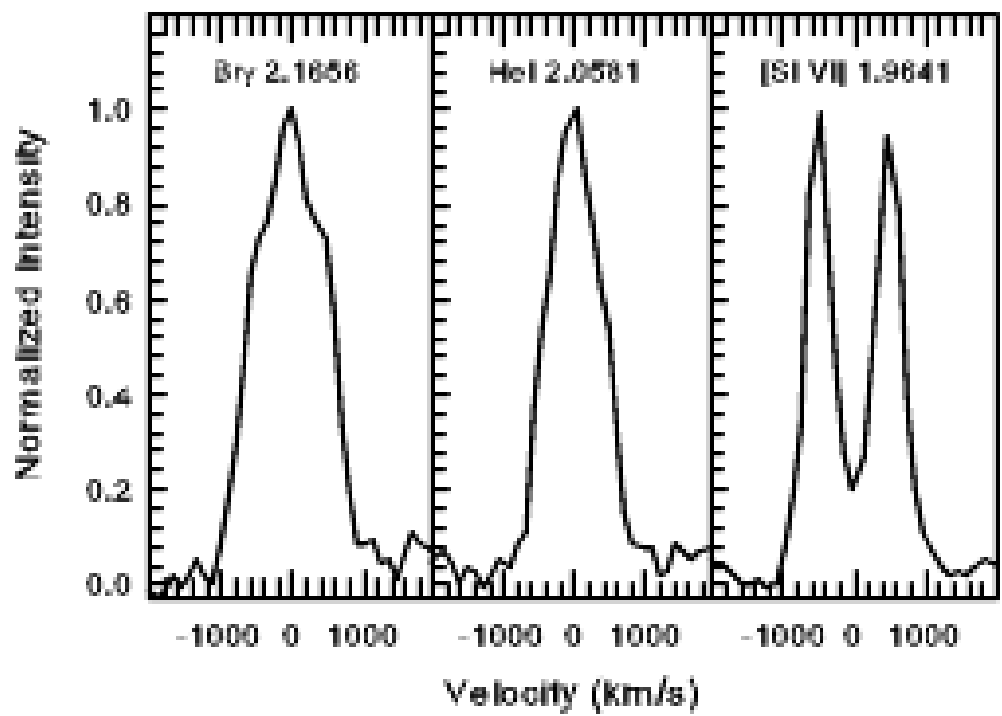

**Figure 16**